# Pattern formation during the evaporation of a colloidal nanoliter drop: a numerical and experimental study

Rajneesh Bhardwaj[a], Xiaohua Fang[b] and Daniel Attinger[a*]

[a]Laboratory for Microscale Transport Phenomena,

Department of Mechanical Engineering,

[b]Langmuir Center for Colloids and Interfaces,

Department of Earth and Environmental Engineering

Columbia University, New York, NY 10027

*Corresponding author. Tel: +1-212-854-2841; fax: +1-212-854-3304;

E-mail address: da2203@columbia.edu (D. Attinger)

**Abstract**

An efficient way to precisely pattern particles on solid surfaces is to dispense and evaporate colloidal drops, as for bioassays. The dried deposits often exhibit complex structures exemplified by the *coffee ring* pattern, where most particles have accumulated at the periphery of the deposit. In this work, the formation of deposits during the drying of nanoliter colloidal drops on a flat substrate is investigated numerically and experimentally. A finite-element numerical model is developed that solves the Navier-Stokes, heat and mass transport equations in a Lagrangian framework. The diffusion of vapor in the atmosphere is solved numerically, providing an exact boundary condition for the evaporative flux at the droplet-air interface. Laplace stresses and thermal Marangoni stresses are accounted for. The particle concentration is tracked by solving a continuum advection-diffusion equation. Wetting line motion and the interaction of the free surface of the drop with the growing deposit are modeled based on criteria on wetting angles. Numerical results for evaporation times and flow field are in very good agreement with published experimental and theoretical results. We also performed transient visualization experiments of water and isopropanol drops loaded with polystyrene microsphere evaporating on respectively glass and polydimethylsiloxane substrates. Measured evaporation times, deposit shape and sizes, and flow fields are in very good agreement with the numerical results. Different flow patterns caused by the competition of Marangoni loops and radial flow are shown to determine the deposit shape to be either a ring-like pattern or a homogeneous bump.

## Nomenclature

| | |
|---|---|
| $a$ | speed of sound [m/s] |
| $c$ | drop liquid vapor concentration [kg/m$^3$] |
| $c_p$ | specific heat [J/kg K] |
| $C$ | dimensionless heat capacity [$\rho c_p/(\rho_L c_{p,L})$] |
| $D_{AB}$ | diffusion coefficient [m$^2$/s] |
| $g$ | gravitational acceleration [9.81 m/s$^2$] |
| $H$ | mean surface curvature [m$^{-1}$], relative humidity [-] |
| $\overline{H}$ | dimensionless mean surface curvature [$Hr_{max,0}$] |
| $j$ | evaporative mass flux [kg/m$^2$ s] |
| $J$ | dimensionless evaporative mass flux [$j/(\rho v_0)$] |
| $k$ | thermal conductivity [W/m-K] |
| $K$ | dimensionless thermal conductivity [$k/k_L$] |
| $L$ | latent heat of vaporization [J/kg] |
| $p$ | pressure [Pa] |
| $P$ | dimensionless pressure [$p/(v_0^2\rho_L)$] |
| $r$ | radial coordinate [m] |
| $R$ | dimensionless radial coordinate [$r/r_{max,0}$] |
| $t$ | time [s] |
| $T$ | temperature [$^o$C] |
| $u$ | radial velocity [m/s] |
| $U$ | dimensionless radial velocity [$u/v_0$] |
| $v$ | axial velocity [m/s] |
| $V$ | dimensionless axial velocity [$v/v_0$], Volume of the drop [m$^3$] |
| $\mathbf{v}$ | velocity vector $\mathbf{v}$ = (u.v) |
| $\mathbf{V}$ | dimensionless velocity vector $\mathbf{V}$ = (U.V) |
| X | Concentration of particles [kg of particles/kg of solution] |
| $z$ | axial coordinate [m] |
| $Z$ | dimensionless axial coordinate [$z/r_{max,0}$] |

**Dimensionless numbers**

*Fr* Froude number [ $v_0^2/(r_{\max,0}g)$ ]

*M* Mach number [$v_0/a$]

*Pr* Prandtl number [$\mu c_{p,L}/k_L$]

*Re* Reynolds number [$\rho v_0 r_{\max,0}/\mu$ ]

*We* Weber number [$\rho v_0^2 r_{\max,0}/\gamma$]

**Greek letters**

$\gamma$ surface energy [$Jm^{-2}$]

$\phi$ wetting angle of the drop [-]

$\mu$ dynamic viscosity [Pa s]

$\theta$ dimensionless temperature [ $\{T_i - \min(T_{1,0}, T_{2,0})\}/|T_{1,0} - T_{2,0}|$ ]

$\rho$ density [$kg/m^3$]

σ stress [Pa]

τ dimensionless time [ $tv_0/r_{\max,0}$ ]

**Subscripts**

0 initial

1 droplet

2 substrate

CL wetting line

f final

int liquid-gas interface

*L* liquid

max maximum value

r radial direction

rec receding, depinning

ref reference value

vap vapor

*z* axial direction

∞ ambient

# 1 Introduction

In biology, the evaporation of colloidal drops is used for depositing or organizing biological materials such as proteins and DNA [1-7]. Colloidal deposition and crystallization [8-14] can also be used to manufacture micro- and nanowires [15, 16], nanocrystals [17] and explosive crystalline layers [18]. Figure 1 shows that the patterns left by an evaporating colloidal drop can exhibit a ring-like structure [19] or more complex features such as a network of polygons [20], hexagonal arrays [17] or a uniform deposit [21, 22]. The pattern variety echoes the complex, coupled and multi-scale transport phenomena occurring during the evaporation of a colloidal drop: fluid dynamics is transient and can be influenced by the impact and associated interfacial deformation by Marangoni stresses, by wetting and the shrinking free surface; heat transfer occurs by convection inside the drop and conduction in the substrate, with a latent heat contribution at the evaporating free surface; mass transfer occurs through diffusion of liquid vapor in the atmosphere, advection-diffusion of particles in the drop and short-range forces between the particles and the substrate.

In 1997, Deegan *et al.* explained the ring deposit formation [19, 23, 24] by a strong radial flow carrying particles towards the pinned wetting line, where evaporative flux is the highest due to the wedge geometry. Their theory [19, 23, 24] was based on the lubrication approximation. For thin drops with spherical cap shapes they provided an analytical expression for the local evaporative flux. Maenosono *et al.* [25] observed that the growth of a nanoparticle ring occurs in two stages, a ring buildup stage with a pinned contact line, followed by a stage where the wetting line recedes. They predicted the ring growth dynamics analytically and found a reasonable agreement with experiments in terms of ring growth rate and final width. Onoda and Somasundaran [26] showed that the chemical or physical heterogeneities of the substrate influence the final deposition pattern. Also, Marangoni convection inside the drop due to the variation of surface tension along the free surface can affect the deposit formation and the flow fields, as shown by Truskett and Stebe [20]. In 2006, Hu and Larson [22] observed that thermal Marangoni convection induces preferential deposition at the center of the droplet rather than at the periphery, for the case of microliter octane drops evaporating on glass coated with perfluorolauric acid (PLFA). Very recently, Ristenpart *et al.* [27] showed analytically that the ratio between substrate and droplet thermal conductivities controls the direction of Marangoni convection inside an evaporating drop. They also showed experimentally that deposit patterns depend on the Marangoni flow direction. The formation of multiple rings has been investigated by Shmulyovich *et al.* [28] during the drying of 1 μL to 70 μL water droplets containing 0.88 μm and 3.15 μm latex particles. The ring formation occurred as a succession of wetting line pinning, ring formation and depinning events. Deegan observed experimentally [23] that the time at which the wetting line of a colloidal drop depins (defined as depinning time) from the deposit corresponds to about 40% to 80% of the total evaporation time. In a very recent article, Zheng [29] showed theoretically that depinning occurs at about 50% of the total evaporation time, independently of the initial concentration of the particles in the drop.

Recently, numerical studies have been reported to study the evaporation of a sessile drop of *pure* liquid on a solid substrate. Hu and Larson [30, 31] developed a finite element model using the lubrication approximation for the evaporation of a sessile drop on a glass substrate with a pinned wetting line. They assumed that the free surface of the drop remains a spherical cap throughout the evaporation. In an extension of this model [32], the same authors reported that Marangoni convection changes the internal flow pattern from radially outward to counter-clockwise. Also, a detailed numerical model for the evaporation of pure microliter water drops was presented by Ruiz and Black [33]. They solved the Navier-Stokes and energy equations with consideration of thermocapillary stresses. In this model, the wetting line was pinned and the evaporative flux was assumed to depend on the substrate temperature but not on the radial distance. Studies by Mollaret *et al.* [34], Girard *et al.* [35] and Widjaja *et al.* [36] followed a similar approach assuming a pinned wetting line but solved the vapor diffusion equation to obtain an accurate evaporative flux.

Regarding *colloidal* drop evaporation, fewer numerical studies have been reported. Within the framework of the lubrication theory, Fischer [37] studied the advection of particles in an evaporating drop, and numerically found transient and local particle distributions within the droplet. Hu and Larson [22] computed the deposition of PMMA particles based on Brownian dynamics simulations for a microliter octane droplet evaporating on glass. The velocities used in this model were found using the lubrication approximation and neglecting convective heat transfer inside the droplet [32]. Using the commercial CFD package CFD-ACE+, Dyreby *et al.* [38] simulated the particle transport in the early stages of the evaporation of a nanoliter drop. This model assumed that the droplet shape is a spherical cap with pinned contact line throughout the evaporation, and that heat transfer and Marangoni convection can be neglected. Also, Dietzel and Poulikakos [39] simulated the coagulation of particles in a drop heated by a laser beam, a phenomenon that takes place so fast that the mass loss through evaporation is negligible. Finally, in 2008, Widjaja and Harris [40] simulated the transport of particles inside an evaporating droplet by solving a continuum advection-diffusion equation.

Due to the complex, coupled physics involved during colloidal drop evaporation, most theoretical and numerical models reported so far are based on assumptions such as fluid flow with negligible inertia [24, 30, 31, 41], small wetting angle [19, 24], spherical cap shape of the free surface [19, 24, 30, 31, 34, 38, 41, 42], pinned wetting line throughout the evaporation [24, 30-35, 40, 41, 43], negligible heat transfer between the drop and the substrate [19, 23, 24, 30, 40, 43] and negligible Marangoni convection [36, 40, 44]. Also, to the best of our knowledge, no published numerical study has considered the receding of the wetting line during colloidal drop evaporation, or the interaction of the free surface and the growing peripheral deposit –which can involve *depinning* i.e. the separation of the drop from the deposit. Thus a numerical modeling involving all these effects would provide a tool to predict deposit growth and final pattern shapes.

Our paper describes a numerical model, which simulates the evaporation of a colloidal droplet with full consideration of the Navier-Stokes equations, convection and conduction heat transfer equations, Marangoni convection, receding of the wetting line. A detailed treatment is proposed for the interaction of the free surface with the peripheral deposit and eventual depinning. The diffusion of vapor in the atmosphere is solved numerically, providing an exact boundary condition for the evaporative flux at the droplet-air interface. The particle concentration is tracked by solving a continuum advection-diffusion equation. The model is validated against published data and used to simulate the formation of deposits during the evaporation of a nanoliter colloidal drop. Finally, the formation of different deposit patterns obtained in our laboratory is explained by our simulations.

## 2 Numerical Model

The mathematical model developed in this study is based on a finite-element code for droplet impact and heat transfer developed by Poulikakos and co-workers in [45-48]. This model has been extensively validated for studies involving impact and heat transfer of molten metal [45, 49] and water droplets [50]. It is extended here for the case of an evaporating colloidal droplet. The flow inside the droplet is assumed to be laminar and axisymmetric. All equations are expressed in a Lagrangian framework, which provides accurate modeling of free surface deformations and the associated Laplace stresses [51]. Our extension of this model for the evaporation of colloidal drops involves the modeling of evaporative flux at the drop free surface, the modeling of thermocapillary stresses and Marangoni flow, the modeling of the convection and advection of particles inside the drop, and the modeling of the wetting line motion in the presence of agglomerating particles. We also propose a criterion to determine if separation of the free surface from the deposit (i.e. depinning) occurs during ring formation. Finally, we have solved the challenge to handle multiple time scales, which range from nanoseconds for capillary waves to several seconds for an entire evaporation process at ambient temperature. The various components of the modeling are presented hereafter.

## 2.1 Fluid dynamics

The radial and axial components of the momentum equation (Equation 1) are considered along with the continuity equation. An artificial compressibility method is employed to transform the continuity equation into a pressure evolution equation (Equation 2), as in [52, 53], and a small Mach number $M$ = 0.001 is used. Only the dimensionless form is given below, since the full derivation is available in [45].

$$\frac{D\mathbf{V}}{D\tau} - \nabla \cdot \mathbf{T} + \frac{\mathbf{n_z}}{Fr} = 0 \text{ (momentum)} \quad (1)$$

$$\frac{DP}{D\tau} + \frac{\nabla \cdot \mathbf{V}}{M^2} = 0 \text{ (continuity)} \quad (2)$$

In the above equations, $M$, $Fr$, $P$, $\tau$, $\mathbf{V}$ are the respective Mach number, Froude number, the dimensionless pressure, dimensionless time and velocity vector, as per definitions in the nomenclature. The definitions of the dimensionless numbers are based on $r_{\max,0}$ = initial wetted radius of the drop and $v_0$ = 1.0 m/s as the respective reference length and velocity. The unit vector $\mathbf{n} = \mathbf{n}_R + \mathbf{n}_Z$ is normal to any boundary considered; $\mathbf{n}_r$ and $\mathbf{n}_z$ are respectively the radial and axial components of the vector $\mathbf{n}$ normal to the boundary and $\mathbf{T}$ is the stress tensor:

$$\mathbf{T} = \begin{bmatrix} \overline{\sigma_{RR}} & \overline{\sigma_{RZ}} & 0 \\ \overline{\sigma_{RZ}} & \overline{\sigma_{ZZ}} & 0 \\ 0 & 0 & \overline{\sigma_{\theta\theta}} \end{bmatrix} \qquad \mathbf{3}$$

The dimensionless stress tensor terms are:

$$\overline{\sigma}_{RR} = -P + \frac{2}{\mathrm{Re}}\frac{\partial U}{\partial R};\overline{\sigma}_{\theta\theta} = -P + \frac{2}{\mathrm{Re}}\frac{U}{R};$$
$$\overline{\sigma}_{RZ} = \overline{\sigma}_{ZR} = \frac{1}{\mathrm{Re}}\left(\frac{\partial U}{\partial Z} + \frac{\partial V}{\partial R}\right);\overline{\sigma}_{ZZ} = -P + \frac{2}{\mathrm{Re}}\frac{\partial V}{\partial Z} \qquad \mathbf{4}$$

where Re is the Reynolds number defined in the nomenclature. It is worth mentioning that due to the Lagrangian formulation the mesh moves with the fluid velocity and advective terms do not appear in the momentum conservation equations, although they are physically considered.

The boundary conditions for the fluid dynamics are described in Figure 2.
Symmetry along the z-axis implies:

$$\text{At } R = 0;\ \ U = 0; \frac{\partial V}{\partial R} = 0 \qquad \mathbf{5}$$

The no-slip and no-penetration boundary conditions are applied at the $z = 0$ plane, except at the wetting line:

$$\text{At } Z = 0\,; R \neq R_{CL}; U = 0; V = 0 \qquad \mathbf{6}$$

Boundary conditions at the wetting line ($R_{\mathrm{CL}}$, 0) are discussed in section 2.5 because they also depend on the evaporation model.

At the free surface, a stress balance is considered including forces due to surface tension, viscous stresses and thermocapillary stresses [39, 54, 55]:

$$(\mathbf{T})^T \cdot \mathbf{n} = -2\frac{\overline{H}}{We}\mathbf{n} + \frac{\nabla\gamma}{\rho v_0^2 r_{\max,0}} \qquad \mathbf{7}$$

where We is the Weber number defined in the nomenclature. In the above equation, $\overline{H}$ is the dimensionless free surface curvature. The temperature dependence of the surface tension $\gamma$ is assumed to be linear as $\gamma = \gamma_0 + (\partial\gamma / \partial T)(T - T_{1,0})$ where subscript '1, 0' represents the initial drop temperature. The Weber number is modified to take into account the variation of surface tension with temperature:

$$\frac{1}{We} = \frac{1}{We_0}\left(1 + \frac{\partial\gamma}{\partial T}\frac{(T - T_{1,0})}{\gamma_0}\right) \qquad \mathbf{8}$$

The initial conditions are that the initial shape of the drop is a sessile spherical cap with a slight overpressure due to Laplace stresses [56]:

$$U = 0; V = 0; P = \frac{8\sin\phi_0}{We} \qquad \mathbf{9}$$

where $\phi_0$ is the initial wetting angle of the drop.

### 2.2 Heat transfer

The energy equation is solved in both the droplet and the substrate, according to the formulation in [45, 47, 48]. Thermal convection and radiation heat transfer from the droplet free surface are neglected. A perfect thermal contact between the drop and the substrate is assumed. The dimensionless energy equations for the drop ($i$ = 1) and the substrate ($i$ = 2) are given by:

$$C_i \frac{D\theta_i}{D\tau} - \frac{1}{\Pr \mathrm{Re}}\nabla^2\theta_i = 0 \qquad \mathbf{10}$$

where $C_i$ is the dimensionless heat capacity, and $Pr$ is the Prandtl number. The symbol $\theta_i$ is the dimensionless temperature defined as:

$$\theta_i = \frac{T_i - \min(T_{1,0}, T_{2,0})}{\left|T_{1,0} - T_{2,0}\right|} \qquad \textbf{11}$$

where $T_{1,0}$ and $T_{2,0}$ are the initial dimensional temperature of the drop and the substrate respectively. In the isothermal case, a reference value of $0^{o}$C replaces $T_{2,0}$ in the above equation. Note that this heat transfer model accounts for both conduction and convection in the drop because of the Lagrangian approach.

The boundary conditions for the heat transfer are also given in Figure 2. At the substrate horizontal surface and at the symmetry axis, we have:

$$\frac{\partial \theta_i}{\partial R}\mathbf{n}_r + \frac{\partial \theta_i}{\partial Z}\mathbf{n}_z = 0 \text{ at } R = 0 \text{ or at } R > R_{\max}, Z = 0 \qquad \textbf{12}$$

Along the side and bottom of the substrate (shown as BC and CD respectively in Figure 2), a constant temperature boundary condition is applied: $\theta_2(R, Z, \tau) = 1$. The substrate is a square of 4×4 non-dimensional size which corresponds to a dimensional value of about 1mm×1mm.

The initial conditions for the droplet and substrate are:

$$\theta_1(r,z,0) = 0\,;\; \theta_2(r,z,0) = 1 \text{ for } T_{1,0} < T_{2,0} \text{ (heated substrate case)} \qquad \textbf{13}$$

$$\theta_1(r,z,0) = 1\,;\; \theta_2(r,z,0) = 1 \text{ for } T_{1,0} = T_{2,0} \text{ (isothermal substrate)}$$

## 2.3 Evaporation model

Since the evaporation model is part of the novel material presented in this paper, it is fully derived below. The evaporative mass flux *j* at the free surface of the evaporating drop is obtained by solving the diffusion equation for the vapor concentration *c* [kg/m$^3$] [30, 31, 41]:

$$\frac{\partial c}{\partial t} = D_{AB}\nabla^2 c \qquad \textbf{14}$$

$D_{AB}$ is the vapor-phase diffusivity of the droplet fluid in air (for water drop, $D_{AB}$ = 2.6e-5 m$^2$/s at 25°C [57]). Importantly, the time required for the vapor concentration to adjust to changes of the droplet shape and surface temperature is on the order of $r_{\max,0}^2 / D_{AB}$ [30], which is 3 orders of magnitude smaller than the total evaporation time of the nanoliter drops considered in this study. Therefore the vapor concentration evolves in a quasi-steady manner with respect to the drop evaporation and we neglect the first term of the above equation.

The boundary conditions for Equation 14 are also described in Figure 2:

$\frac{\partial c}{\partial z} = 0$ at $r > r_{\max}, z = 0$; (No penetration of the vapor concentration into substrate surface)

$\frac{\partial c}{\partial r} = 0$ at $z > h_{\max}, r = 0$; (Axisymmetry) **15**

$c = Hc_{\infty}$ at $r = \infty, z = \infty$ ; (Constant vapor concentration in the far-field )

$c = c_{vap}$ at the drop-air interface (Constant vapor concentration at the drop-air interface)

where $r_{\max}$ and $z_{\max}$ are the wetted radius and the height of the drop, respectively. In the above equation, $c_{\text{vap}}$ is the saturated concentration [kg/m$^3$] of the vapor and $c_{\infty} = Hc_{\text{vap}}$ is the concentration in the far-field corresponding to the ambient relative humidity $H$. The saturated concentrations are fitted from the data in [58, 59]:

$$c_{vap} = [9.99e - 7T^3 - 6.94e - 5T^2 + 3.20e - 3T - 2.87e - 2] \text{ (for water)} \qquad \mathbf{16}$$

$$c_{vap} = [5.12e - 6T^3 - 2.51e - 4T^2 + 1.12e - 2T - 5.463e - 2] \text{ (for isopropanol)}$$

where $T$ is the temperature in °C. We found sufficient to consider the far field ($r = \infty, z = \infty$) at $r = 20r_{\max,0}, z = 20r_{\max,0}$ (Figure 2).

The evaporative mass flux $j$ at the free surface can be expressed as in [31, 41]:

$$\mathbf{j}(r,T) = D_{AB}(T)\left[\frac{\partial c}{\partial r}\mathbf{n}_r + \frac{\partial c}{\partial z}\mathbf{n}_z\right]\Bigg|_{\text{int}} \qquad \mathbf{17}$$

At the free surface of the drop, the hydrodynamic and thermodynamic vapor-liquid jump conditions [34, 54] are applied:

$$\mathbf{j}\cdot\mathbf{n} = \rho(1 - \boldsymbol{X}_{\text{int}})(\mathbf{v} - \mathbf{v}_{\text{int}})\cdot\mathbf{n} \quad \textbf{18}$$

$$\mathrm{jL} = -\mathrm{k}\nabla\mathbf{T}\cdot\mathbf{n} \quad \textbf{19}$$

where $\mathbf{v}$ is velocity of the liquid *at* the free surface and $\mathbf{v}_{\text{int}}$ denotes the velocity *of* the free surface. The symbol $\mathbf{n}$ is the outward normal unit vector at the liquid-air interface and $\rho$ is the density of the drop liquid. Also, $X_{\text{int}}$ is the volume concentration of the particles at the free surface and $L$ is the latent heat of evaporation of the liquid [J/kg].

## 2.4 Particle transport

The governing equation for the particles transport is given by [1, 6, 60] :

$$\frac{DX}{Dt} = D_{PL}\nabla^2 X \quad \textbf{20}$$

where $X$ is the concentration of the particles [kg of particles/kg of solution] and $D_{PL}$ is the diffusion coefficient of the particles in the drop liquid [$m^2$/s]. The values of the diffusion coefficient for 100 micron particles in water, 1 micron particles in water and 1 micron particles in isopropanol are calculated using the Stokes-Einstein equation [61] as 4e-12, 4e-13 and 2e-13 $m^2$/s, respectively. Boundary conditions are given as in Figure 2:

At $r = 0$, $\partial X/\partial r = 0$ (axisymmetry)

At $z = 0$ $\partial X/\partial z = 0$ (no penetration of the particles into the substrate)

We neglect any attraction force between the substrate and the colloidal particles because in all the systems considered here, the substrate and colloidal particles repel each other, having zeta potentials of the same polarity.

The respective non-dimensional expressions for the hydrodynamic and thermodynamic vapor-liquid jump conditions (Eqs 18 and 19) are: $\mathbf{J}\cdot\mathbf{n} = (\mathbf{V} - \mathbf{V}_{\text{int}})\cdot\mathbf{n}$ and $J\,\mathrm{Re}\,\mathrm{Pr}\,Ja = -K_1\nabla\theta\cdot\mathbf{n}$ where $J = j/\rho_l v_0$ and the Jacob number $Ja = L(T)/[c_{p,l}|T_{1,0} - T_{2,0}|]$.

In the isothermal case, a reference value of 0°C replaces $T_{2,0}$ in the denominator. The non-dimensional particle transport equation can be formulated as $DX/D\tau = D_{PL}^{*}\nabla^{2}X$ where $D_{PL}^{*} = D_{PL}/(r_{\max,0}v_{0})$

## 2.5 Boundary conditions at the wetting line and criterion for depinning

As shown in Figure 3 in our model, the drop is initially sessile and the wetting line is allowed to recede only if the following two conditions are simultaneously verified:

- the wetting angle is smaller than the receding wetting angle ($\phi_{rec}$), and
- the concentration of particles at the interface ($X_{CL}$) is lower than the maximum packing concentration (0.7 for spherical particles [62]).

Otherwise, the wetting line does not move: it is pinned. In the *pinned* case, the Laplace stresses (Equation 7) are extrapolated to the wetting line (natural boundary condition) and the wetting line velocity $\mathbf{v}\cdot\mathbf{n} = \mathbf{j}\cdot\mathbf{n}/\rho(1-X_{\text{int}})$ is an essential boundary condition obtained from eq. 18, with $\mathbf{v}_{\text{int}} = 0$ at the pinned point.

In the case where the wetting line is *receding*, the Laplace stresses (Equation 7) are extrapolated to the wetting line (natural boundary condition) and the wetting line velocity $\mathbf{v}$ is obtained as an essential boundary condition from eq. 18, $\mathbf{v}\cdot\mathbf{n} = \mathbf{v}_{\mathbf{int}}\cdot\mathbf{n} + \mathbf{j}\cdot\mathbf{n}/\rho(1-X_{\text{int}})$ where $\mathbf{v}_{\text{int}}$ is calculated using the fact that the wetting angle is constant ($\phi = \phi_{rec}$).

Figure 3 explains the depinning or receding of the wetting line during the evaporation of a *pure liquid drop.* According to the Young-Dupre equation [29, 56], depinning starts when the wetting angle of the drop $\phi$ becomes equals to $\phi_{\text{rec}}$ (critical value of the receding angle, usually determined experimentally). As shown in Figure 3, during the early stages of the drop evaporation, the wetted radius remains constant and the wetting angle $\phi$ decreases until $\phi_{\text{rec}}$ is reached at $t = t_2$. After that, depinning occurs and the wetted radius decreases while the wetting angle remains constant at its receding value ($\phi = \phi_{\text{rec}}$), until complete evaporation of the drop.

In the case of a *colloidal drop*, a particle concentration $X$ corresponding to the maximum packing ($X = 0.7$) might be reached at the wetting line so that a ring starts growing as in Figure 4. As the wetting line becomes cluttered by the assembling particles, we propose in Figure 4 to describe the interaction of the free surface of the drops with the growing deposit as follows. First, we consider that once a closely-packed ring starts forming, the wetting line shifts from its natural location (along the edge of the drop and on the substrate, at point C in Figure 4) to a different location: the intersection of the ring and the free surface of the drop (point I in Figure 4). In Figure 4, the boundary of the growing ring is determined by the iso-concentration front of 0.7 shown as a thick red line inside the drop in Figure 4. To propose a mechanism for depinning of the free surface from the growing particle ring, we will use the receding wetting angle $\phi_{rec}$ (left column of Figure 4) of the pure liquid drop on a substrate made of closely-packed colloidal particles. This angle $\phi_{rec}$ can be obtained experimentally and is needed to explain the interaction of the colloidal drop free surface with the growing ring (right column of Figure 4), during evaporation on a smooth solid substrate. The top frames in Figure 4a show how depinning occurs at point (I). At $t = t_1$, the wetting angle $\phi$ at point (I) is greater than $\phi_{rec}$. As evaporation proceeds, the wetting angle $\phi$ decreases until reaching at $t = t_2$ the depinning value $\phi = \phi_{rec}$. At that instant, depinning occurs and the depinning line slides along the 0.7 concentration front with $\phi = \phi_{rec}$. The bottom frames in Figure 4b show that a smaller wetting angle ($\phi'_{rec}$) for the (pure liquid)-air-particles system leads to a situation where the colloidal ring forms without depinning because the wetting angle $\phi$ at point I is always larger than $\phi'_{rec}$. Obviously, our model relies on measuring the wetting angle $\phi_{rec}$ for the pure liquid drop on a substrate made of closely-packed colloidal particle material, the value of which might be influenced by the surface roughness, on the order of the particle diameter. Note that at the interface between the drop liquid and the ring, a horizontal velocity boundary condition is applied with a magnitude matched to the evaporation rate at the surface of the ring. Also, mesh elements of which all the nodes have reached 0.7 are removed from the fluid dynamics calculation, according to the scheme defined in [63].

## 2.6 Numerical Scheme

The mathematical model presented in the previous sections is solved using the Galerkin finite element method [64] with a fluid dynamics scheme from Bach and Hassager [65], as described in [45]. That scheme is however specifically suited for transient drop impact problems, with a time step constrained to the smallest time scale of the problem, which happens to be the oscillation of parasitic capillary waves on the droplet free surface. This time scale of $\Delta x/a$ is about 15 ns, with $\Delta x$ the grid size (~ 2 μm) and $a$ is the speed of sound. However, the entire evaporation process takes about 2 minutes for a 20 nL water drop at ambient temperature: this would involve 8 billion time steps and a prohibitive computing time. To resolve this issue, we have developed a two-step temporal integration scheme. First, a converged, instantaneous solution of the fluid flow and the evaporation flux (Equations 1-9 and 14-18) is obtained using a short time step ($\Delta\tau_{short}$, order of 15 ns) with the Bach and Hassager scheme [65], and we account for evaporation with equation 18 to determine the shrinking of the free surface. Second, we perform a quasi-steady Lagrangian integration with a long time-step, about two orders of magnitude larger ($\Delta\tau_{long}$, order of 1 μs), using the converged velocities from the previous, short time step. The computation proceeds by alternating between the two time steps. This two-step approach is justified because the fluid flow and the evaporation flux can be considered steady for the duration of the long time step because $\Delta\tau_{long} << \tau_{evaporation}$. To compensate for the artificial loss of particles due to the shrinking of the free surface, a source term $S = Xj\,\delta A\,/\,\rho\delta V$ is added to the diffusion equation (eq. 20) for the elements touching to the free surface of the drop, where $X$ is the particles concentration, $\rho$ is the density of the drop liquid, $j$ is the evaporative mass flux, and $\delta A$ and $\delta V$ are respectively the free surface area and volume of the element. Boundary conditions at the wetting line and on the ring are applied as described in section 2.5.

The typical computational domain is shown in Figure 2 and involves a spatial discretization of 650 nodes in the drop of linear, triangular elements, refined near the wetting line, to ensure grid and time step independency. The influence of grid size and time step have been extensively checked for heat transfer and fluid dynamics case in [45] and in [66] for the evaporation of pure liquids. The typical computational time was about two days on an Intel dual core processor.

## 2.7 Thermophysical properties and dimensionless numbers

The thermophysical properties used in the simulations for water, isopropanol, glass and polydimethylsiloxane (PDMS) are given in Table 1, with values taken from Refs [57, 58, 67, 68]. The temperature dependence of the dynamic viscosity $\mu$, vapor-phase diffusivity of the droplet fluid in air $D_{AB}$ and latent heat $L$ is taken into account. All quantities are defined in SI units, as shown in the nomenclature. For this purpose, the following curves are fitted on the data in [57, 59]:

For water [57] $\mu(T) = (-0.0026T^3 + 0.5874T^2 - 47.598T + 1763.4)1e-6$ Pa-s

$$D_{AB}(T) = (2.5e-4)\exp(-\frac{684.15}{(T+273.15)}) \text{ m}^2\text{/s}$$

$$L(T) = (-2.5314T + 2508.6)1e3 \text{ J/kg} \qquad \textbf{21}$$

For isopropanol [57, 59]

$$\mu(T) = (-1.80e-5T^3 + 2.0672e-3T^2 - 1.4817e-1T + 4.619)1e-3 \text{ Pa-s}$$

$$D_{AB}(T) = (1.0e-4)\exp(-\frac{690.0}{(T+273.15)}) \text{ m}^2\text{/s}$$

$$L(T) = (-1.8286T + 813.43)1e3 \text{ J/kg}$$

where the temperature $T$ is in °C.

## 2.8 Code validation

While several aspects of our numerical modeling have been extensively validated, such as the fluid dynamics [69, 70] and associated heat transfer [45], we evaluate below the ability of the present modeling to simulate the evaporation of colloidal drops. First, we apply our model to the evaporation of pure liquids and compare the results with published data. Then, we test the conservation of mass and energy during colloidal drop evaporation. Finally, we compare the simulated transient drop volume and accumulated mass in the ring with published analytical models of evaporating colloidal drops.

### 2.8.1 Evaporation of a pure liquid drop with constant receding angle

Many published experiments describe the drying of millimeter-size drops [71-74] of pure liquids, but fewer describe the drying of smaller drops, which is the focus of the present study. In [75], Erbil *et al.* recorded the evaporation of a 480 nL toluene drop on a Teflon substrate: the process started with a wetted radius of 895 μm and proceeded with a constant receding wetting angle of 44$^o$. The ambient temperature and relative humidity were respectively 22.1°C and 54%. Our simulation results for this case are presented in Figure 5. Thermophysical properties such as $\partial\gamma/\partial T$ = -1.19e-4 Nm$^{-1}$K$^{-1}$ are taken from [76]. Figure 5a compares the numerical and experimental transient evolution of the liquid volume and drop radius. The agreement between experiments and simulations is excellent. Also, the non-linear evolution of the wetted radius during the receding phase is compared to a simple analytical model available in the literature [77-83], which can be described as follows: during a receding phase with constant wetting angle $\phi_{rec}$, the wetted radius for the spherical cap evolves as $r_{\max} \propto (t_f - t)^a$, where $t_f$ is the total evaporation time. The exponent $a$ depends weakly on the nature of the fluid, being slightly less than 0.5 for alkanes [77, 81, 82] but close to 0.6 for water [23, 24, 77]. Based on this analysis, Figure 5a shows that an analytical fit $R_{\max} = 0.077(168 - t)^{0.5}$ is indeed in excellent agreement with numerical results. Numerical results are plotted in Figure 5b as dimensionless isotherms, streamlines and velocity magnitudes at $t$ = 30 s. In the simulations, the fluid is warmer near the wetting line than at the top of the drop. As a consequence, a surface tension gradient along the free surface generates an upward Marangoni flow, which combines with the radial flow needed to replenish the evaporating region near the contact line and results in a counter-clockwise flow loop visible in Figure 5b.

### 2.8.2 Conservation of thermal energy and mass of particles

The conservation of thermal energy by the numerical simulation has been verified for the typical case of a 20 nL pure water drop with 45° initial wetting angle evaporating on an insulated glass substrate initially at 100°C, as plotted in Figure 6. Energy is conserved over time if $E_{1,0} + E_{2,0} = E_1(t) + E_2(t) + E_{loss}(t)$ where $E_{1,0}$ and $E_{2,0}$ are the initial thermal energy of the drop and the substrate, respectively, and $E_1$ and $E_2$ are the thermal energy of the drop and the substrate, each expressed as $E_i(t) = \int_{V_i} \rho_i C_{p,i} (T_i(t) - T_{ref}) dV$. The thermal loss due to phase change at the evaporating free surface is $E_{loss}(t) = \int_0^t \int_{A_{\text{int}}} j(r)(L(r) + C_p (T_{\text{int}}(r) - T_{ref})) dA dt$, with $T_{\text{ref}}$ = 0°C. Figure 6 plots the drop volume and the relative value of the thermal energy $(E_1 + E_2 + E_{loss}) \times 100/(E_{1,0} + E_{2,0})$ as a function of time. Our results show that thermal energy is conserved within 1%. Also, the mass conservation is verified by plotting the mass of the colloidal particles at any time with respect to the initial particle loading (1% v/v concentration of 100 nm particles). We observe that the mass of the particles is also conserved within 1% and conclude that the modeling conserves mass and energy very well.

### 2.8.3 Evaporation of a pinned colloidal drop

In Figure 7, the temporal evolution of the particle mass in the ring obtained through the numerical model is compared with the analytical expression by Deegan *et al.* [24], an expression valid for thin drops evaporating at ambient temperature:

$$m = m_0[1-(1-\frac{t}{t_f})^{\frac{1+\lambda}{2}}]^{\frac{2}{1+\lambda}} \qquad 22$$

In Equation 22, $m_0$ is the initial mass of particles in the drop, $t_f$ is the total evaporation time and $\lambda=(\pi - 2\phi_0)/(2\pi - 2\phi_0)$ a function of the initial contact angle. We consider two drops, a 20 nL and a 2 nL water drops evaporating on glass with a low value for the initial wetting angle (12°) and a pinned contact line, because Deegan's analysis is based on lubrication theory and valid only for thin drops with pinned contact line. The drop and the substrate are at ambient temperature (25°C). The comparisons of the evolution of the non-dimensional mass of the ring given by analytical theory and by the simulation are shown in Figure 7. The trend of the ring mass given by the numerical results and the analytical theory are in good agreement, although the numerical growth of the ring mass for the 2nL drop appears delayed.

Figure 7 also plots the temporal evolution of the drop volume, which can be found from an analytical theory by Popov [41]:

$$V = \frac{\pi r_{\max,0}^3 \phi_0}{4}(1-\frac{t}{t_f}) \qquad 23$$

In the above equation [41, 84], the drying time of the drop is proportional to the square of the value of the initial wetted radius:

$$t_f = \frac{\pi \rho_l r_{\max,0}^2 \phi_0}{16 D_{AB}(c_{\text{int}} - c_\infty)} \qquad 24$$

In Equation 24, $\rho_l$ is the density of liquid [kg/m$^3$], $r_{\max,0}$ is the initial wetted radius of the drop, $D_{AB}$ is the vapor-phase diffusivity of the droplet fluid in air [m$^2$/s], $c_{\text{int}}$ and $c_\infty$ are the vapor concentrations [kg/m$^3$] respectively near the interface and in the ambient air.

Figure 7 shows a good agreement between our simulations and the analytical formulae above for both the mass in the ring and the drop volume evolution. However, in the later stages of the evaporation, the numerical volume decreases slightly faster than predicted analytically. A possible explanation is that our simulations were made with 5% volume of particles while the correlations [41, 84] and [24] do not specifically consider the concentration of particles. Also, results in Figure 7 only show the first two third of the evaporation process, corresponding to a total computation time of about two days.

## 3 Experimental details

Nanoliter water and isopropanol drops were spotted on glass and PDMS substrates, using a 375 μm diameter stainless steel pin, as routinely done to prepare bioarrays [85]. Using the same standard pin, deposited water drops were smaller (3-5 nL) than isopropanol drops (30-40 nL) probably because of higher wettability of steel by the solvent [85]. Evaporating droplets were visualized from the side using a digital camera (Pixelink, PLA 741, 1.3 Megapixel up to 100 frames/second) and an Optem long-distance zoom objective. Typical time and spatial resolution were respectively 4 frames per second and 1 μm per pixel. We also used an Olympus IX-71 inverted microscope to qualitatively assess the structure of the flow inside the drop: the motion of fluorescent particles (1-micron or 100 nm polystyrene particles from Duke scientific Inc, CA) was recorded at frame rates of 25 frames per second, and resulted in figures such as Figure 16 and the associated movie [86]. After evaporation, profiles of the deposits were measured using an Atomic Force Microscope (Digital Instruments Dimension 3000 with Si3N4 tip) in contact mode, as in Figure 13. While this method worked well for deposits on glass, measurements on PDMS substrates failed: the AFM tips broke possibly because of the sticky PDMS. Therefore, the profile of deposits on PDMS was measured by the following imprinting method [87]: First, the PDMS substrate (Dow corning Inc) was coated with a thick PDMS slab having a different ratio of base and curing agent, this to facilitate subsequent removal. After curing, the thick PDMS slab was removed and washed to remove particles from the ring imprint. Then, epoxy (Loctite Inc, quick set # 81502) was poured and cured on the PDMS slab. The positive imprint of the deposit was finally measured on the epoxy with a surface profilometer (Dektak3), as in Figure 18.

## 4 Results and Discussions

First we describe two purely numerical results related to the Marangoni flow during evaporation. The first result relates the direction of the Marangoni flow loop to the ratio of substrate and droplet thermal conductivities, as per an analytical criterion (section 4.1). The second numerical result describes the effect of the intensity of Marangoni convection on the deposit shape (section 4.2). Then, we compare numerical and experimental results for two specific colloidal evaporating cases that form different patterns:

(a) Evaporation of a 3.7 nL water drop on a glass substrate with 1% volume fraction polystyrene particles having 100 nm or 1 μm diameter. The result is a peripheral ring (section 4.3).

(b) Evaporation of a 38 nL isopropanol drop on a PDMS substrate with 0.1% volume fraction polystyrene particles having 1 μm diameter. The result is a homogeneous central bump. (section 4.4).

Simulations were performed at ambient temperatures using the same parameters as in the experiments, with thermophysical properties and simulation parameters as in Table 1 and 2.

### 4.1 Direction of the Marangoni loop

Using asymptotic analysis and experiments, Ristenpart *et al.* [27] showed that the direction of the Marangoni loop is controlled by the ratio $k_2/k_1$ of substrate vs. liquid thermal conductivities as shown in Figure 8(right column). An interesting question is therefore to determine if our numerical loops also obey the criterion in [27], and results are shown in Figure 8. The idea behind the criterion in [27] is as follows. For small values of $k_2/k_1$ (such as 0.6 in Figure 8), the heat supplied to the wetting line (where evaporation is the highest) comes preferentially from the liquid drop. This creates a thermal gradient along the free surface that points towards the top of the drop, and results in a clockwise Marangoni flow, as in Figure 8(left column, D). Note that we name the rotation direction of the flow in reference to our figures, where the streamlines are plotted right from the axial symmetry axis. In reality, a 'clockwise' Marangoni flow is a flow where streamlines ascend along the axis of symmetry to the top of the drop, radially descend along the free surface towards the wetting line and come back to the center along the substrate. A 'counterclockwise' flow is in the exact opposite direction. For large values of $k_2/k_1$, (such as 1.8 or 10 as in Figure 8) the heat supplied to the wetting line comes preferentially from the substrate resulting in an inversion of the thermal gradient along the drop free surface so that the Marangoni loop becomes counterclockwise. In Figure 8(left) we consider a system of 30 nL isopropanol drop on PDMS substrate with an initial wetted radius and wetting angle of respectively 437 μm and 32°, For this system, Ristenpart *et al.* [27] found the critical ratio for $k_2/k_1$ (when the Marangoni loop switches direction) to be 1.57. Isotherms and streamlines in Figure 8(left) show that this switching is successfully reproduced by our code, as shown by the different rotation direction for $k_2/k_1$ of respectively 1.77 and 0.6. Note that a similar reasoning can explain the counterclockwise nature of the flow in the toluene drop reported in the section 2.8.1, for which $k_2/k_1 = 2.05$ is larger than a critical value of about 1.6 (see Figure 5b, right). Interestingly the numerical simulations predict a more complex flow pattern for $k_2/k_1 = 1.2$ with two loops, which might occur to the proximity of the transition region. In Figure 8(left), the thermal conductivity of the substrate was the only parameter varied, all other drop and substrate properties being kept constant.

## 4.2 Influence of Marangoni flow on pattern formation

Figure 9 investigates how the magnitude of the thermocapillary flow controls the regimes of ring formation or uniform deposit pattern, for the case of a 38 nL isopropanol drop with 1% of 1 μm polystyrene particles evaporating on a polydimethylsiloxane (PDMS) substrate. Two simulations are performed where the value of $\partial\gamma/\partial T$ is arbitrarily varied by one order of magnitude. The initial value of the drop wetted radius and wetting angle are respectively 437 micrometer and 32°. The drop and the substrate are initially at ambient temperature (25°C). The simulation with the stronger thermocapillary effect (on the right column of Figure 9) shows a vigorous Marangoni loop that prevents particles to accumulate at the wetting line. The simulation with the weaker thermocapillary effect (on the left column of Figure 9) shows a weaker Marangoni loop that does not fully prevent the radial flow to carry particles towards the wetting line, where evaporation is the highest.  Therefore at about 16 s, the wetting line pins and starts forming a ring visible as an inset in Figure 9(left column). Figure 9(right column) however does not exhibit any ring. This numerical result shows that by reducing the intensity of Marangoni convection inside the drop, it is possible to change the final deposit pattern.

## 4.3 Evaporation of a 3.7 nL water drop on glass

Figure 10 and the associated movie [88] describes our numerical results for the evaporation of a 3.7 nL colloidal water drop on a glass substrate at ambient temperature, with an initial 1% v/v concentration of 100 nm polystyrene particles. Time goes from top to bottom. It shows streamlines and velocity amplitude (right) as well as isoconcentrations of particles (left, noted X). The initial value of the drop wetted radius and wetting angle obtained from experiments are respectively 231 μm and 22°. As in simulations from other groups involving water [32, 89], Marangoni stresses are neglected because the high surface tension of water favors the deposition of contaminants at the water/air interface, a process that inhibits Marangoni convection. The parameters used in the simulation are given in Table 2, with a receding angle ($\phi_{rec}$) around 2-4° [30]. As a consequence of this low receding angle, the wetting line is pinned during the whole evaporation of the drop as described in section 2.5. Inspection of our results in Figure 10 reveals that at 8 s and 10 s, the internal fluid flow develops as radially outward, advecting particles towards the wetting line. This flow pattern arises because of pinning and maximum evaporation flux at the wetting line, as explained by Deegan *et al.* [19] and Hu and Larson [31]. In relation to that we plot in Figure 11 the evaporation flux, obtained numerically at $t$ = 0 s, with respect to the radial distance along the substrate. Our flux results agrees extremely well with the correlation of Hu and Larson [31], with both approaches showing the strongest evaporation at the wetting line.

In Figure 10 the advected particles accumulate near the wetting line and start forming a ring as soon as the particle concentration reaches the 0.7 limit corresponding to maximum particle packing, visible in red color from $t$ = 8 s. A magnified view of the growing polystyrene particles ring is visible at the bottom of Figure 10, from $t$ = 11.7 s. At that point, the drop has flattened into a film and we have skewed the vertical scale. Since the receding wetting angle of water on polystyrene is about $80^o$ [90], depinning of the drop free surface from the ring might occur according to the modeling in Figure 4a, and described in section 2.5. However, our simulation in Figure 10 assumes a receding angle of water on polystyrene of 45° to account for the roughness of the ring. This prevents depinning as in Figure 4b. As a result, depinning does not occur and the ring is formed by the drying of the central film. While it is difficult to assess how the depinning angle varies with the particle diameter and associated ring roughness, we show in Figure 13 that for the colloidal system considered in Figure 10 numerical simulations where depinning did not occur produced a ring shape closer to the experimental ring shape.

Figure 12 compares the experimentally measured and simulated evolution of the volume, wetted radius and wetting angle during the evaporation of the same colloidal water drop on glass. In both numerical and experimental results the wetted radius is approximately constant during the evaporation of the drop (from $t$ = 0 to 11.7 s). This is explained by the pinning of the wetting line, as described above. The numerical results for the evolution of the drop volume shows a linear decrease from $t$ = 0 to 11.7 s, and a similar trend can be noticed in the experiment. A linear decrease in the evolution of the wetting angle can also be noticed in the numerical as well as the experimental result. The final evaporation time of the drop is around 12 s, slightly faster in the experiment than in the simulation.

Figure 13 compares the profiles of the ring measured by Atomic force microscopy (AFM) with the simulation result. Three AFM measurements of deposits generated under the same conditions as in Figure 10 are quite similar. They show that the ring has the profile of a dome, about 1.7 μm high and 35 μm wide. Two profiles are obtained numerically, each corresponding to a different value of the depinning angle for water on the ring material, as described in Figure 4 and above. The numerical profile corresponding to the lower value of the depinning angle ($\phi_{rec}$ = 45°, as in simulation in Figure 10) appears to match the experiments quite well, better than the profile with the larger depinning angle ($\phi_{rec}$ = 85°) where depinning seems to have occurred too abruptly. While both numerical simulations match the outer slope of the ring very well, the inner slope suggests that the model with $\phi_{rec}$ = 45° is more appropriate, and this might be explained by the fact that the roughness due to the particles assembling in the ring reduces the value of the depinning angle of the water drop.

An interesting question regarding the numerical modeling presented in this paper is to determine how large the particles can be before the continuum modeling for particle transport becomes inadequate. Therefore, numerical and experimental results are shown in Figure 14 for the same case as Figure 13, except that 1 μm particles are used (10 times larger than the 100nm particles in Figure 13). AFM measurements in Figure 14 show a ring with a rectangular cross-section corresponding exactly to one monolayer of particles. While the ring width is comparable to the width obtained with 100nm particles (Figure 13), the height does not compare well with the simulation results in Figure 14, for the reason postulated directly above.

## 4.4 Evaporation of a 38 nL isopropanol drop on PDMS

Figure 15 and the associated movie [91] describe our numerical results for a situation where the outcome is not a ring shape but a homogeneous deposit. The system corresponds to a 38 nL isopropanol drop evaporating on a PDMS substrate at ambient temperature. The drop is initially loaded with 0.1% concentration of 1 μm polystyrene particles. Figure 15 shows streamlines and velocity amplitude (right) as well as isoconcentrations of particles (left, noted X). The initial value of the drop wetted radius and wetting angle are obtained from our experiments as 437 μm and 32°, respectively. Marangoni stresses are taken into account with $\partial\gamma/\partial T$ = -7.89e-5 $Nm^{-1}K^{-1}$ and the physical parameters used in the simulation are given in Table 2. Since the ratio of substrate and liquid thermal conductivities $k_2/k_1 = 1.77 > 1.57$, we expect the Marangoni loop to turn in counterclockwise direction, according to the criterion in section 4.1. This is verified in Figure 15, for $t$ = 3 to 12 s. The receding angle ($\phi_{rec}$) obtained experimentally is around 27°, slightly smaller than the initial wetting angle, so that the wetting line is initially pinned and the wetting angle gradually decreases. At $t$ = 2 s, the wetting angle hits its receding value and the wetting line recedes with a constant receding wetting angle of 27°, according to the mechanism explained in section 2.5. Figure 15 also shows that evaporation increases the concentration in the vicinity of the free surface. This phenomenon, together with the Marangoni convection drags particles along the free surface towards a stagnation point visible on the top surface of the drop from $t$ = 3 to 12 s. As time evolves, more and more particles are advected to the top of the drop where they accumulate ($t$ = 12 s).

This accumulation of particles on the top of the drop agrees well with experiments in Figure 16 and the associated movie [86], using the same conditions as in Figure 15. We visualize the flow inside the drop using fluorescence microscopy: by carefully adjusting the height of the focal plane we were able to see the radial component of the velocities at different levels in the drop, as done in [27]. The blurred image of the off-focus bead ($t$ = 3.8 s) indicates that the circulation is counter-clockwise, in agreement with our numerical result in Figure 15 and Figure 16b. An interesting fact we noticed from our experiments is that once the drop was spotted on the substrate, the particles moved in a chaotic manner for the first 15% of the evaporation time (3 s), until a slow and axisymmetric flow took place. This phenomenon is consistent with observations made for a 2 microliter isopropanol drop on PDMS substrate by Ristenpart *et al.* [27]. We believe the chaos comes from oscillations that are caused by the deposition process from the pin.

However Ristenpart *et al*. [27] speculated that this initial chaotic flow is due to Benard-Marangoni instability.

Figure 17 compares the experimental and numerical evolution of the volume, wetted radius and wetting angle for the same 38 nL colloidal isopropanol evaporating on a PDMS substrate. Experiments show that the receding of the wetting line starts at 2.5 s. The numerical results also show that a pinned phase precedes a receding phase. Also, simulations show that the wetting angle decreases linearly during the first stage of the evaporation while it is constant during the second stage of the evaporation (from $t$ = 2 to 15 s). The final evaporation time of the drop is 15 s according to the simulations, close to the measured value of 18 s.

The profile of the pattern obtained experimentally and numerically is shown in Figure 18. The large oscillations and poor reproducibility of the measurement probably stem from the indirect stamping process explained in section 3. However the general shape, volume and dimensions of the measured deposits are in good agreement with our simulation results: they show a single central bump that differs radically from the peripheral ring seen in the water-glass case of section 4.3. Our numerical results explain this difference by the change of flow pattern due to Marangoni convection, as described in Figure 15.

# 5 Conclusions

In this work, the formation of deposits during the drying of nanoliter colloidal drops on a flat substrate has been investigated numerically and experimentally. A finite-element numerical model has been developed that solves the Navier-Stokes, heat and mass transport equations in a Lagrangian framework. The diffusion of vapor in the atmosphere is resolved numerically, providing an exact boundary condition for the evaporative flux at the droplet-air interface. Laplace stresses and thermal Marangoni stresses are accounted for. The particle concentration is tracked by solving a continuum advection-diffusion equation. For the first time, the receding of the wetting line and the interaction of the free surface of the drop with the growing deposit have been modeled using wetting angle criteria. Evaporation times and flow field calculated numerically are in very good agreement with published experimental and theoretical results, as well as with in-house experiments. For instance, numerical comparisons with experiments involving water and isopropanol drops loaded with polystyrene microspheres show the importance of Marangoni convection in controlling the internal flow and final deposit pattern, to be either a ring-like pattern or a single central bump. Numerical results are found to be in very good agreement with the measured deposit shapes and sizes.

## 6 Acknowledgements

The authors gratefully acknowledge financial support for this work from the Chemical Transport Systems Division of the US National Science Foundation through grant CTS-0622849. We also thank Howard A. Stone from Harvard University, MA, Ilona Kretzschmar from City College of New York, NY, William Ristenpart from the University of California at Davis, CA, Ponisseril Somasundaran from Columbia University, NY, Matthias Dietzel from Caltech, CA and Yiannis Ventikos from Oxford University, UK for useful discussions.

## 8 Tables

Table 1: Thermophysical properties used in the simulations at 25$^o$C [27, 57, 59, 92, 93]

| Substance | Density [$kgm^{-3}$] | Thermal conductivity [$Wm^{-1}K^{-1}$] | Specific heat [$Jkg^{-1}K^{-1}$] | Viscosity [Pa-s] | Surface energy [$Jm^{-2}$] | Gradient of surface tension with temperature [$Nm^{-1}K^{-1}$] | Latent heat [$Jkg^{-1}$] |
|---|---|---|---|---|---|---|---|
| Water | 997 | 0.607 | 4180 | 9.0e-4 | 7.2e-2 | -1.68e-4 | 2445e3 |
| Isopropanol | 780 | 0.13 | 2560 | 2.04e-3 | 2.09e-2 | -7.89e-5 | 768e3 |
| Glass | 2200 | 1.38 | 740 | - | - | - | - |
| PDMS | 1030 | 0.23 | 1460 | - | - | - | - |

Table 2: Parameters used in the simulations reproducing experiments

| Case | $r_{cap,i}$ μm | $h_{cap}$ μm | $V_d$ nL | $\phi_i$ | $\phi_{rec}$ | $T_{1,0}$ | $T_{2,0}$ | $\gamma = f(T)$ | $\mu = f(T)$ | $H$ |
|---|---|---|---|---|---|---|---|---|---|---|
| Water-glass | 231 | 44 | 3.7 | 22$^o$ | - | 25$^o$C | 25$^o$C | No | Yes | 50% |
| Isopropanol-PDMS | 437 | 125 | 38 | 32$^o$ | 27$^o$ | 25$^o$C | 25$^o$C | Yes | Yes | 48% |

# 9 Figures

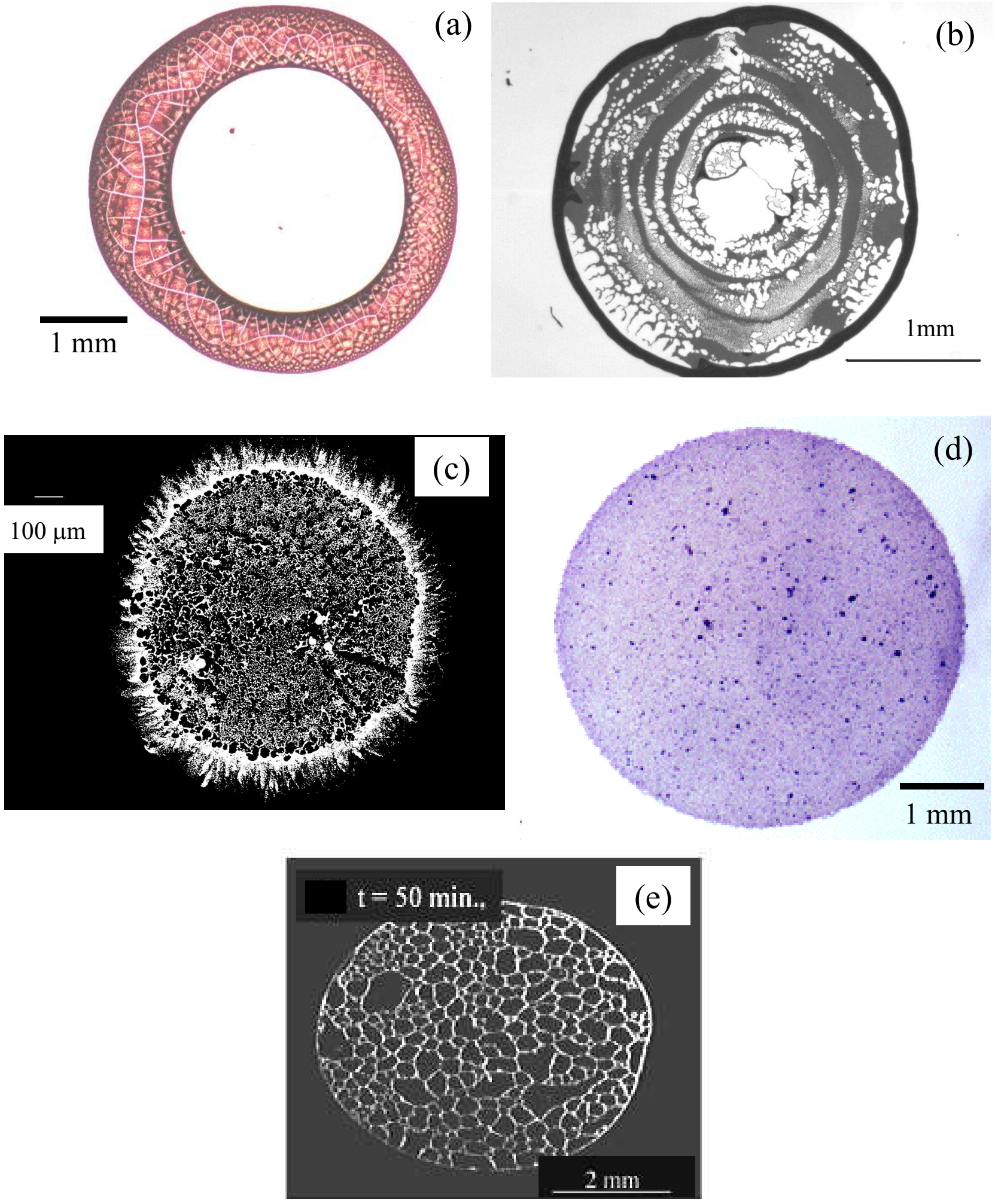

Figure 1: A multiplicity of deposits can be obtained after the drying of a colloidal drop: (a) ring-like pattern from an aqueous drop containing 60 nm polystyrene spheres on titanium substrate [94] (with permission from ACS); (b) multiple rings from a μL water drop containing 1 μm polystyrene microspheres on glass (our work); (c) fingering at wetting line obtained from a μL isopropanol drop with 1μm polystyrene microspheres on glass (our work); (d) uniform deposition pattern of 60 nm hydroxyapatite particles from aqueous drop on titanium disk [95] (with permission from ACS); (e) hexagonal cells from surfactant-laden aqueous drop containing polystyrene microspheres on hydrophobic OctadecylTricholoroSilane (OTS) substrate [20] (with permission from ACS).

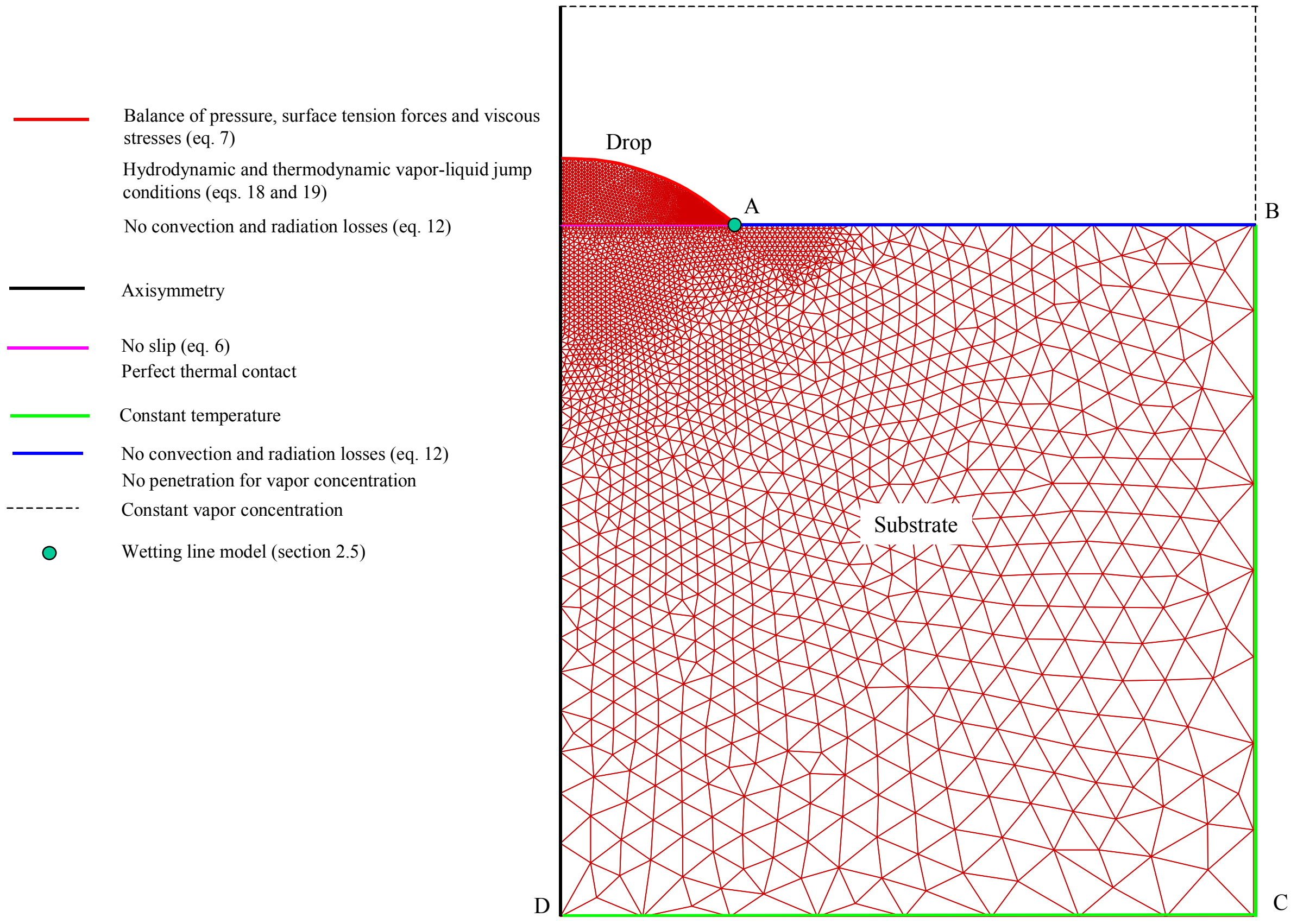

Figure 2:Boundary conditions used in the mathematical model. A typical mesh used in this study.

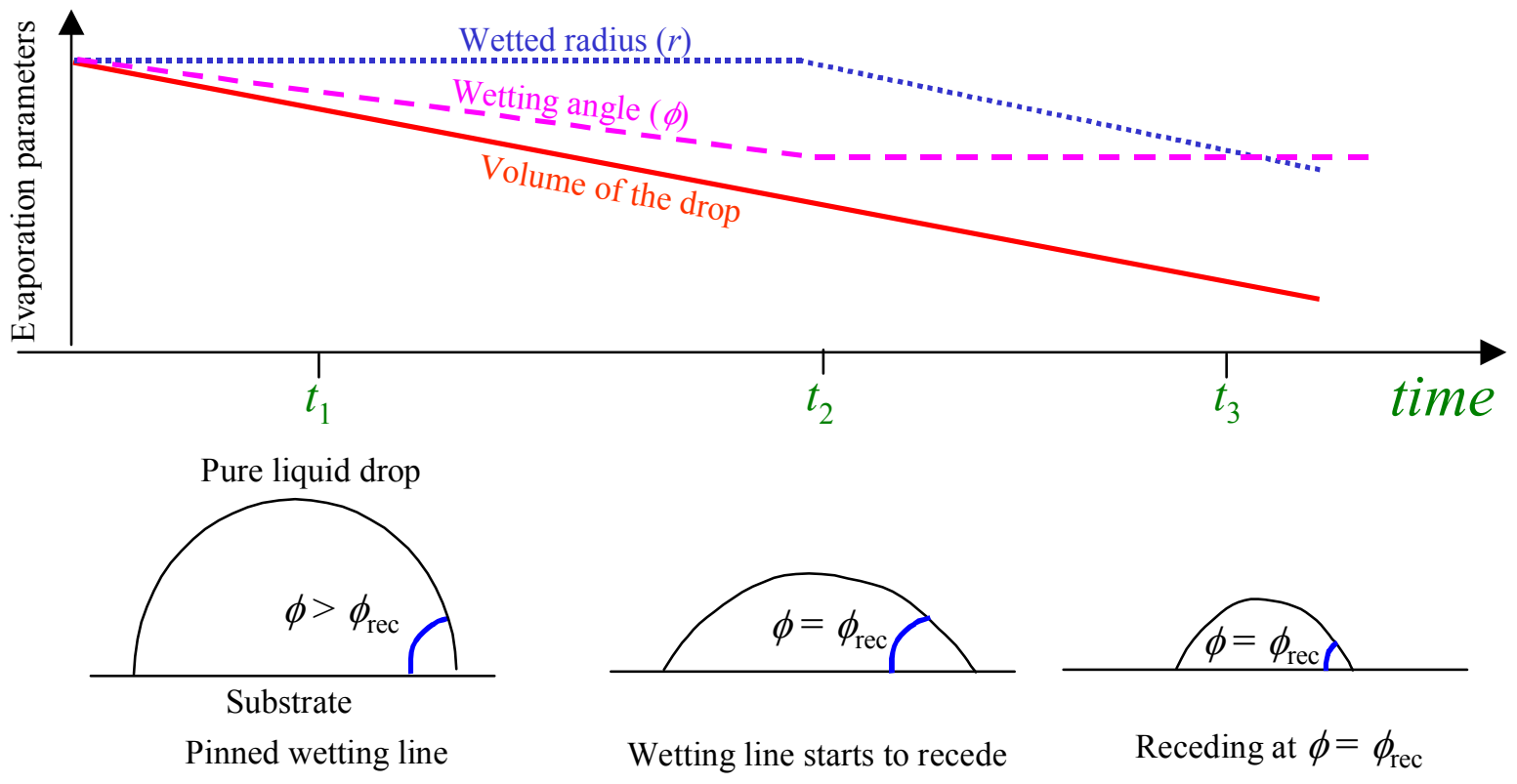

Figure 3: Depinning mechanism for the evaporation of a pure liquid drop. Receding starts when the wetting angle reaches a given receding value $\phi_{rec}$ and then proceeds at constant wetting angle.

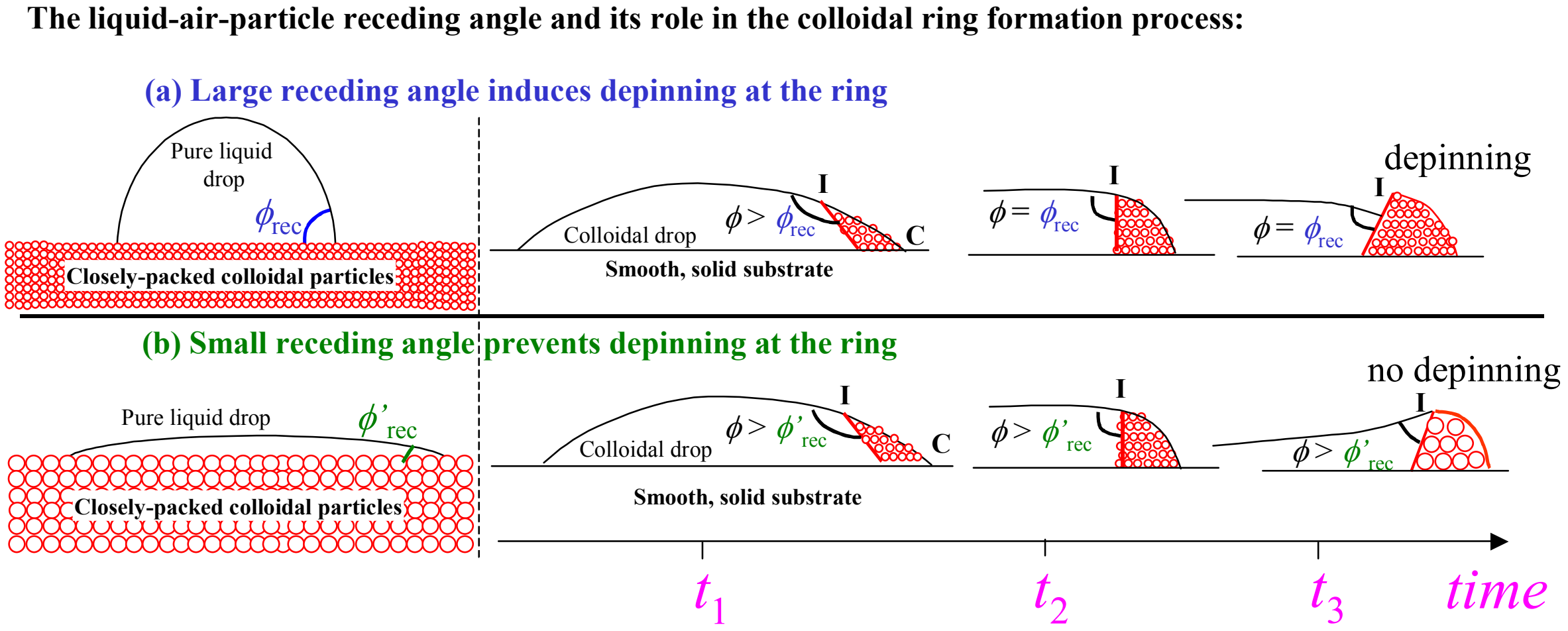

Figure 4: A proposed mechanism for depinning of the free surface from a growing ring made of colloidal particles. We first define a wetting angle $\phi_{rec}$ (left) as the receding angle of the pure liquid drop on a substrate made of closely-packed colloidal particle material (measured on the left side). The frames on the right describe the different system of the colloidal liquid drop, a smooth solid substrate and air. The top frames (a) show that depinning at point (I) occurs during drying of the colloidal drop on a smooth solid substrate when $\phi_{rec}$ is reached at point (I). The bottom frames (b) show that a smaller wetting angle ($\phi'_{rec}$) for the pure liquid-air-particles substrate leads to a situation where the ring forms without depinning.

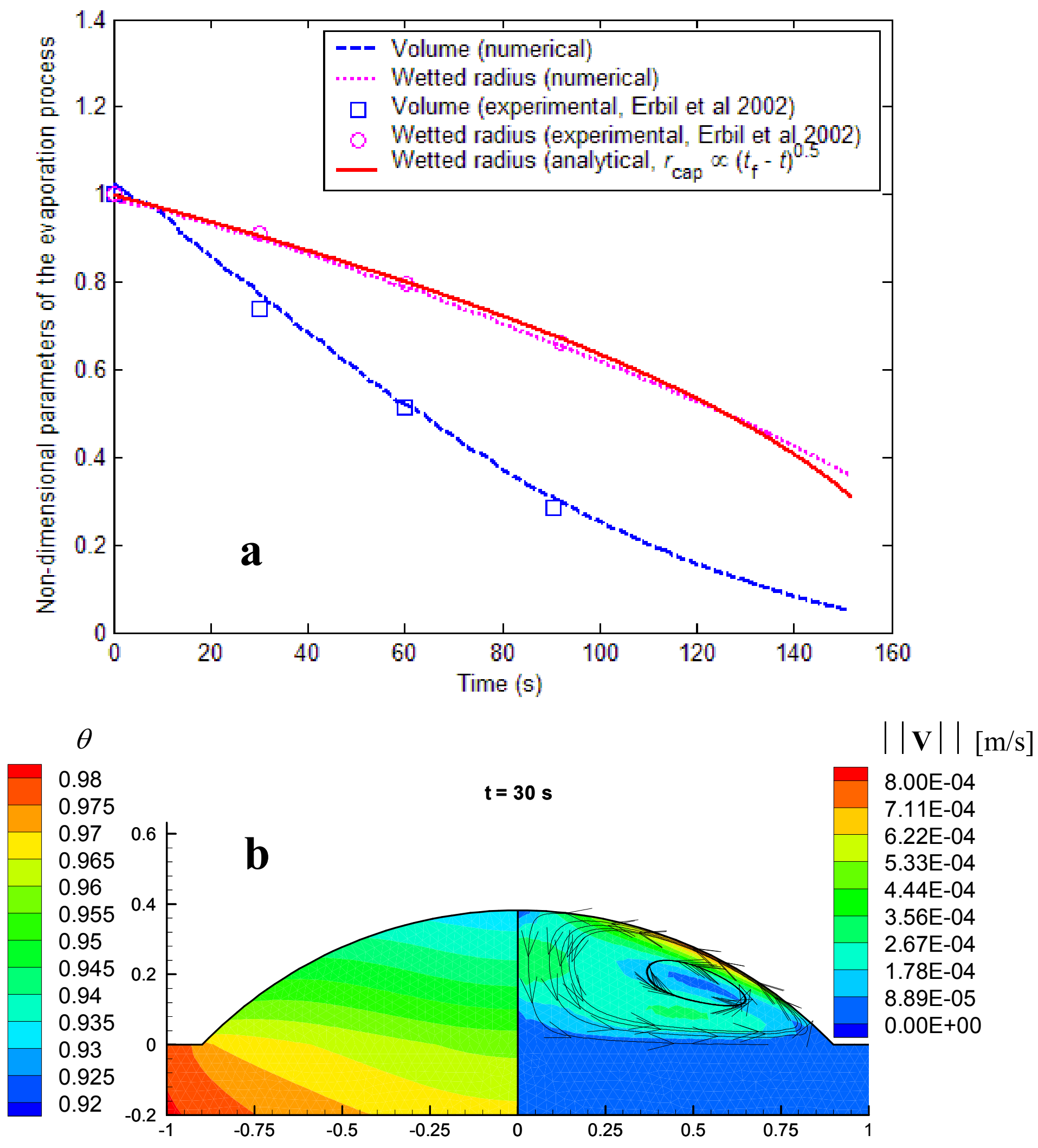

Figure 5: Comparison of results of our numerical code with published data [75] for the case of a pure 480 nL toluene drop evaporating on a Teflon substrate. Initial substrate temperature, initial wetting angle and initial wetted radius of the drop are 22.1°C, 44° and 895 µm, respectively. (a) Comparison of experimental and numerical results for the evolution of volume and wetted radius of the drop, normalized with respect to their values at $t$ = 0; (b) numerical results showing temperature distributions (left) and Marangoni flow loop (right) at $t$ = 30 s.

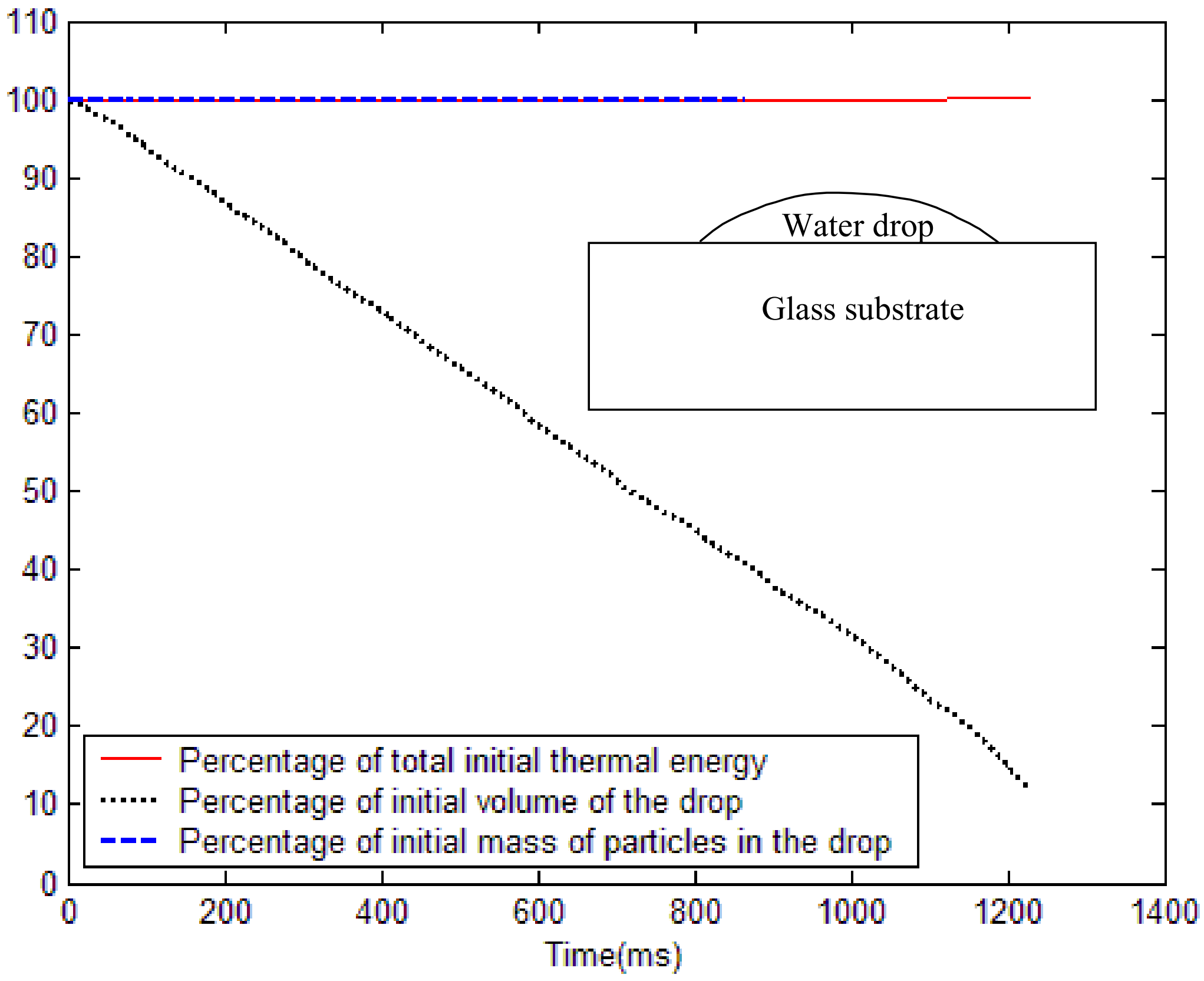

Figure 6: Conservation of thermal energy and particles mass during the evaporation of a 20 nL colloidal water drop on a heated glass surface.

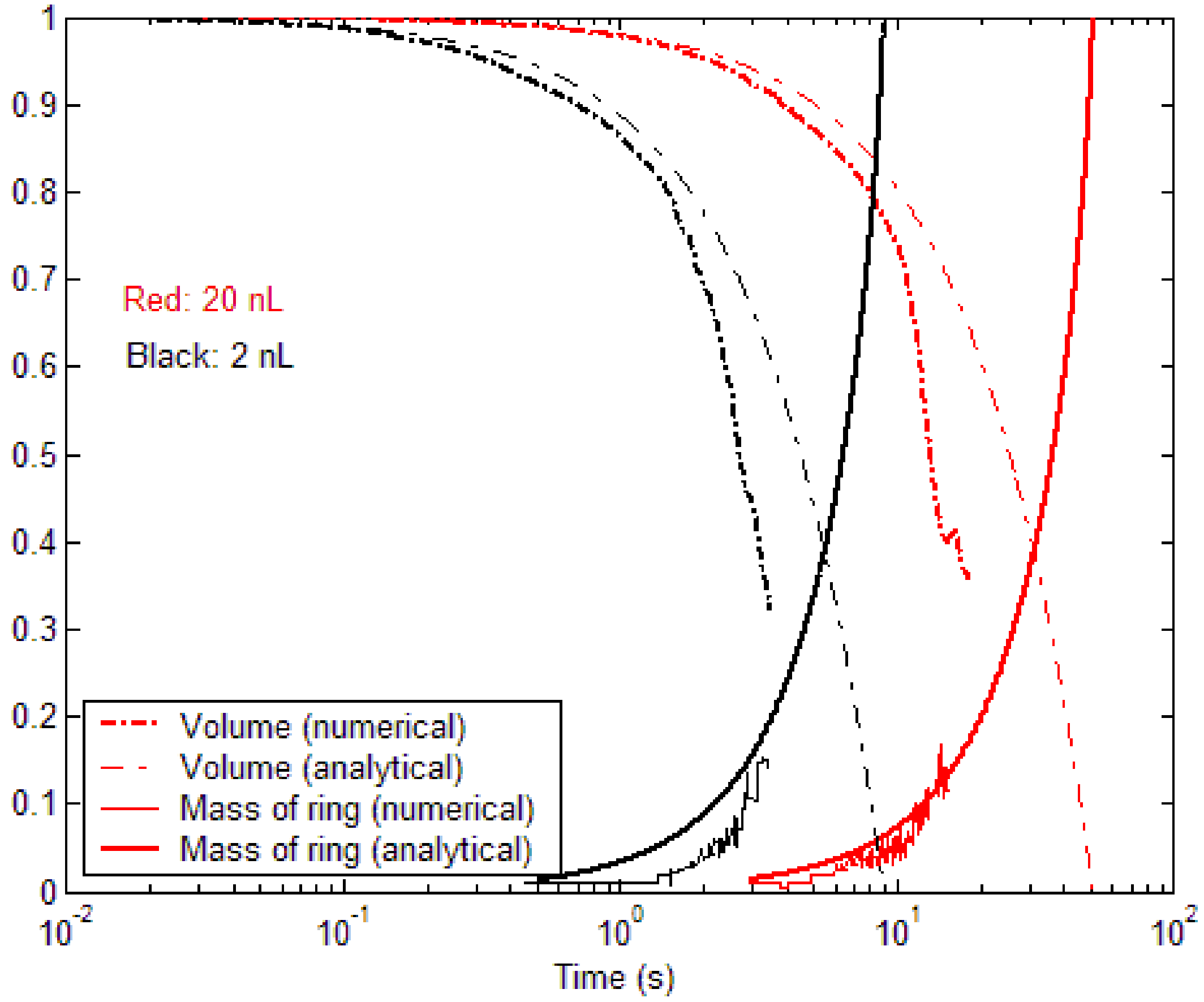

Figure 7: Comparison of our numerical results and analytical results for the evolution of the drop volume [41] and ring mass [24] for a 20 nL and 2 nL water drop evaporating isothermally on a glass substrate with initial wetting angle of 12°.

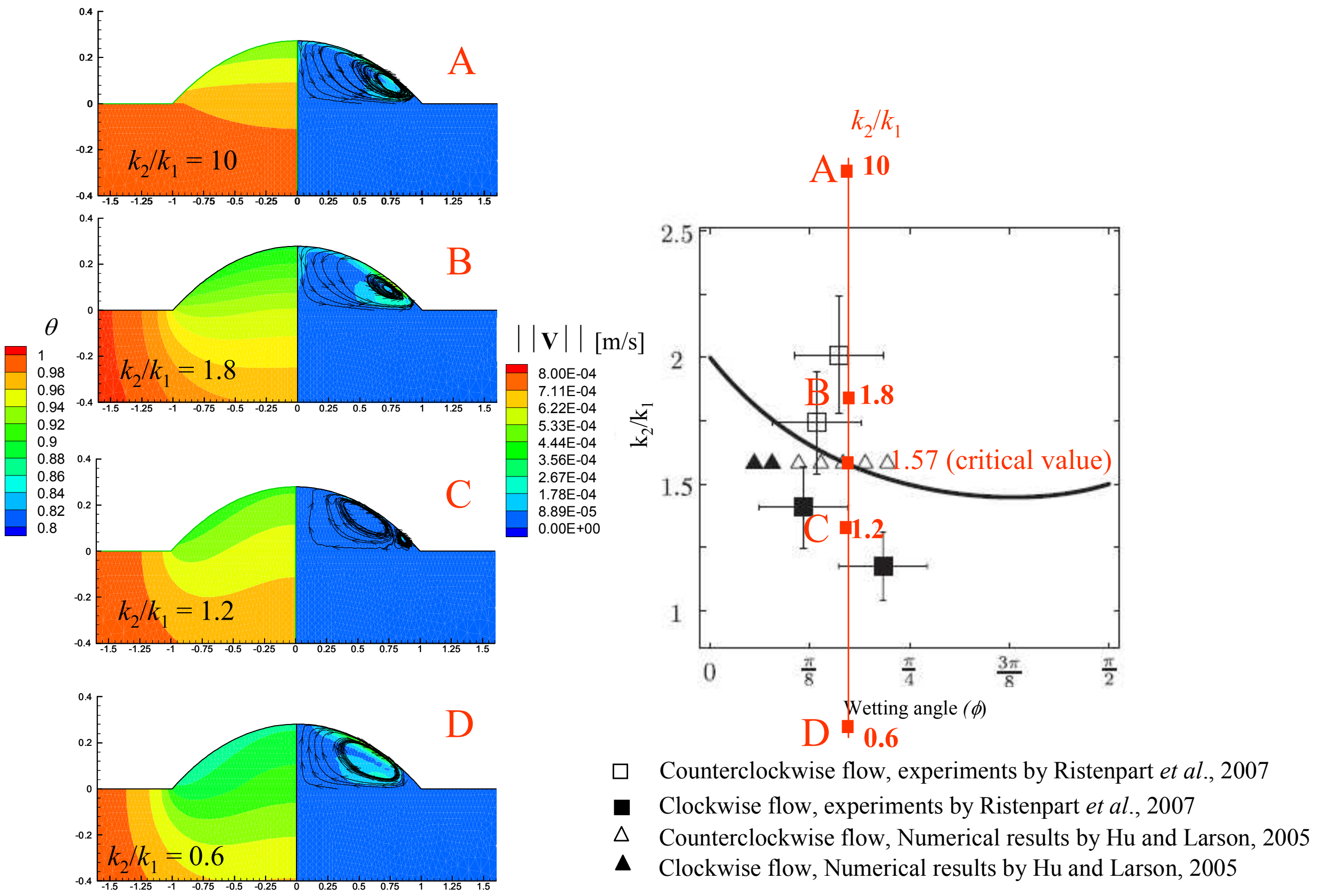

Figure 8: Influence of $k_2/k_1$, the ratio of substrate and liquid thermal conductivities on the direction of the Marangoni loop for the evaporation of a 38 nL isopropanol drop on a PDMS substrate, with initial substrate temperature, wetting angle and wetted radius of respectively 25°C, 32° and 437 µm respectively The present numerical modeling (left column) show temperature contours (left), and streamlines superposed on the velocity amplitude contours (right) shows a transition in the loop direction, which confirms analytical results (right column) from [27] (permission pending to reproduce this chart).

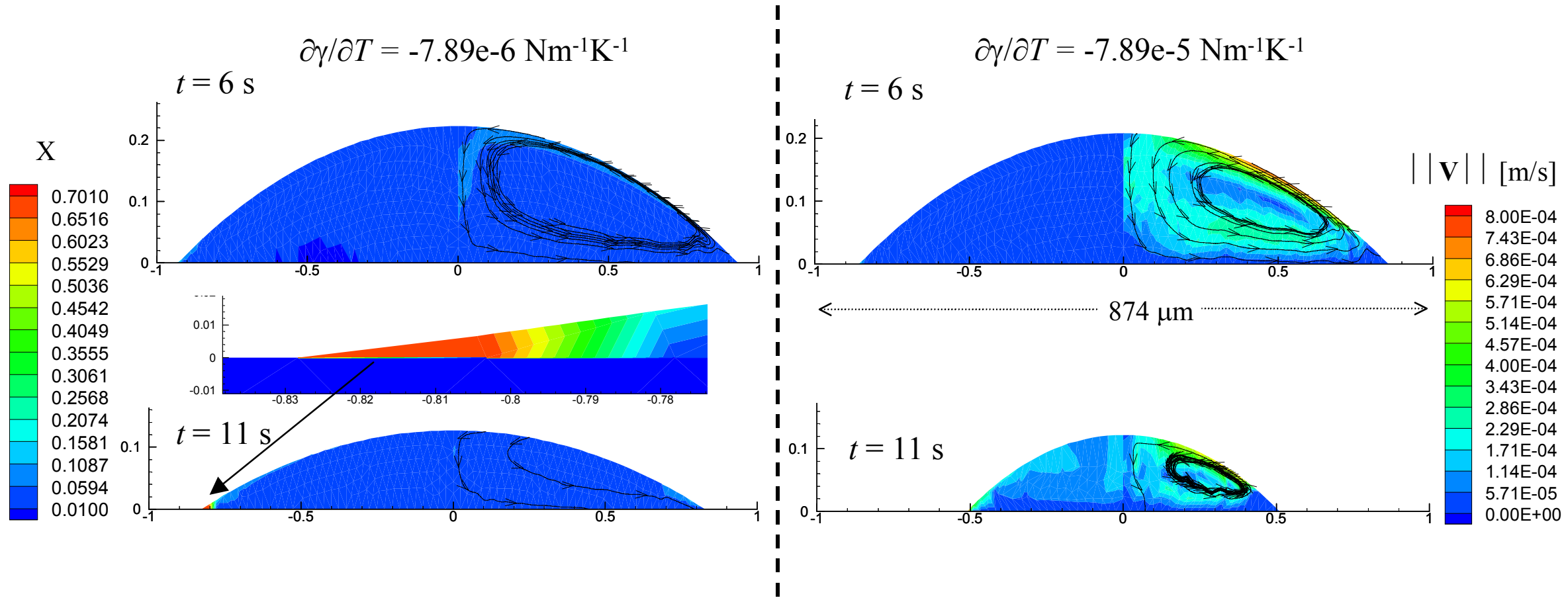

Figure 9: Influence of the intensity of Marangoni convection on pattern formation. Left column shows results for $\partial\gamma/\partial T$ = -7.89e-6 $Nm^{-1}K^{-1}$ while the right column corresponds to $\partial\gamma/\partial T$ = -7.89e-5 $Nm^{-1}K^{-1}$. Isoconcentration of the particles (left) and streamlines superposed to velocity amplitude (right) are shown in each frame inside the drop at different times.

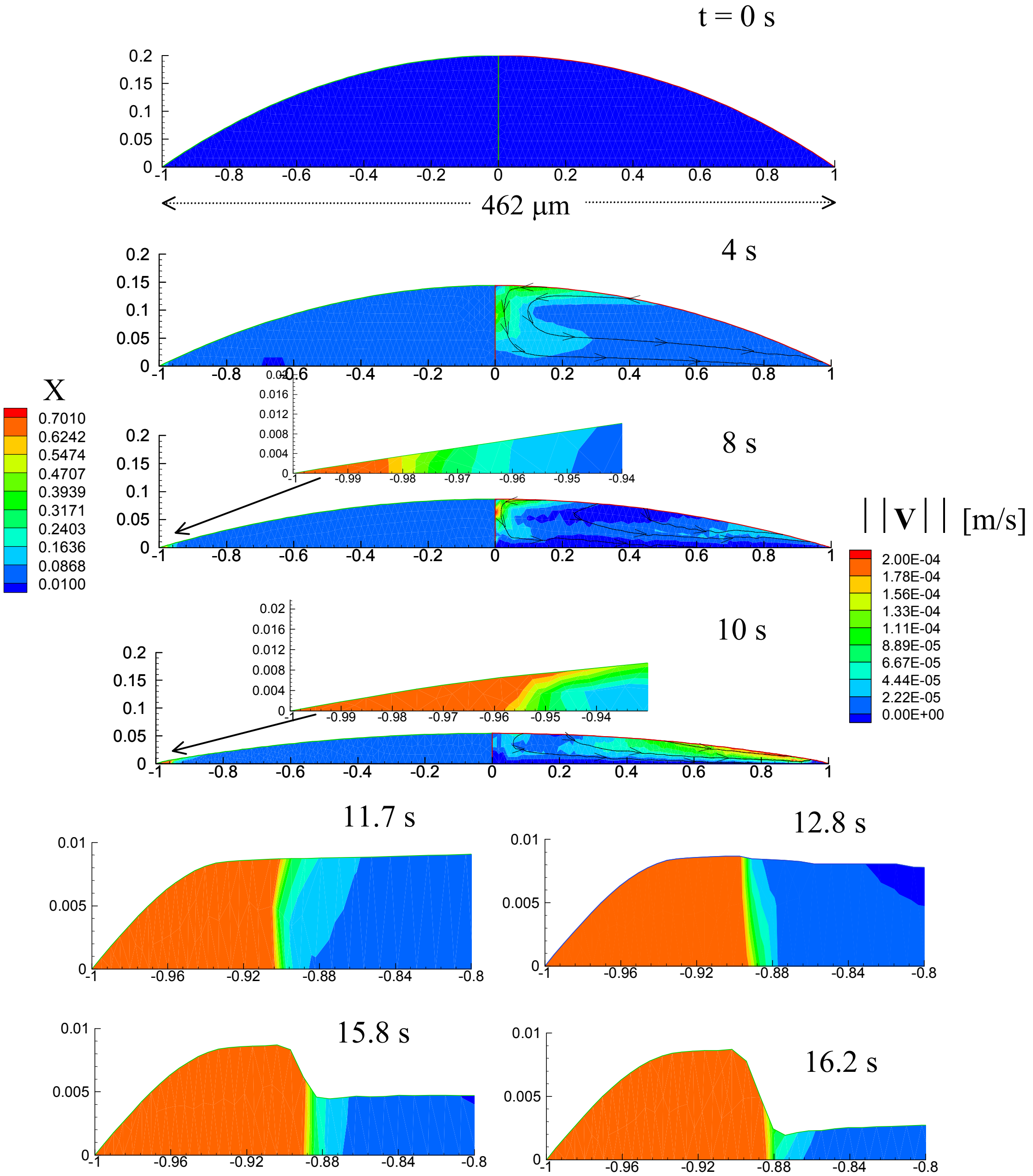

Figure 10: Simulation of the evaporation of a 3.7 nL colloidal water drop on glass with ring formation. Particle concentration contours (left) and streamlines superposed to velocity amplitude (right) are shown. A ring starts forming at the wetting line when the concentration reaches 0.7 (in red, 8 s). Note that the aspect ratio is increased towards the vertical axis from $t$ = 11.7 to 16.2 s. See also the associated movie [88].

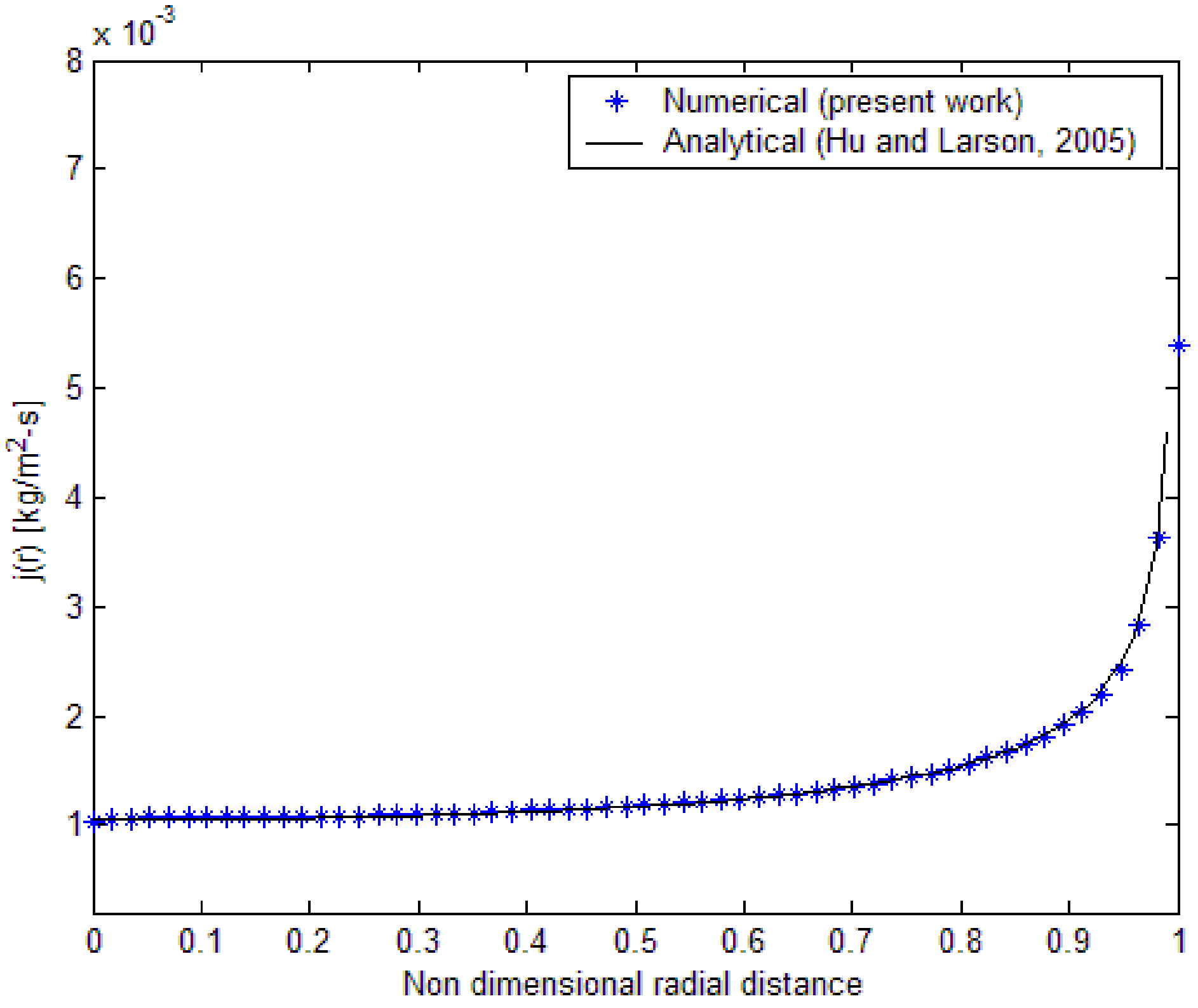

Figure 11: Comparison of numerical and analytical profile of evaporative flux for a pure 3.7 nL water drop on glass at $t$ = 0 s. The drop and the substrate are at ambient temperature (25$^{o}$C). Initial wetting angle and initial wetted radius of drop are 22$^{o}$ and 231 μm, respectively.

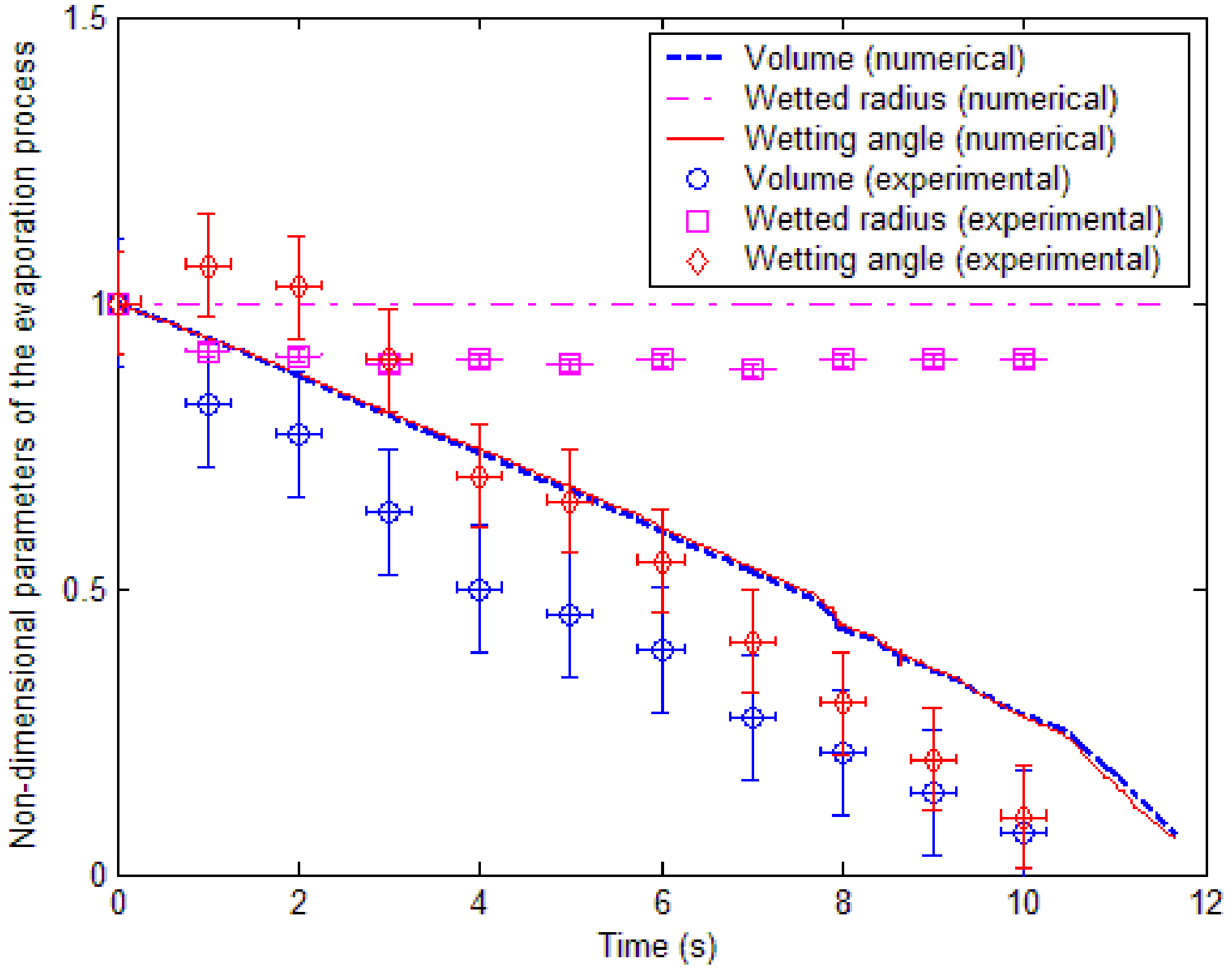

Figure 12: Time evolution of numerical and experimental volume, wetted radius and wetting angle for the evaporation of a 3.7 nL water drop on glass. Initial values of substrate temperature, wetting angle and wetted radius of drop are 25°C, 22° and 231 μm respectively. Comparison of the evolution of volume, wetted radius and wetting angle is shown. These values are normalized with respect to their initial values (at *t* = 0). Error bars are also shown for experimental results.

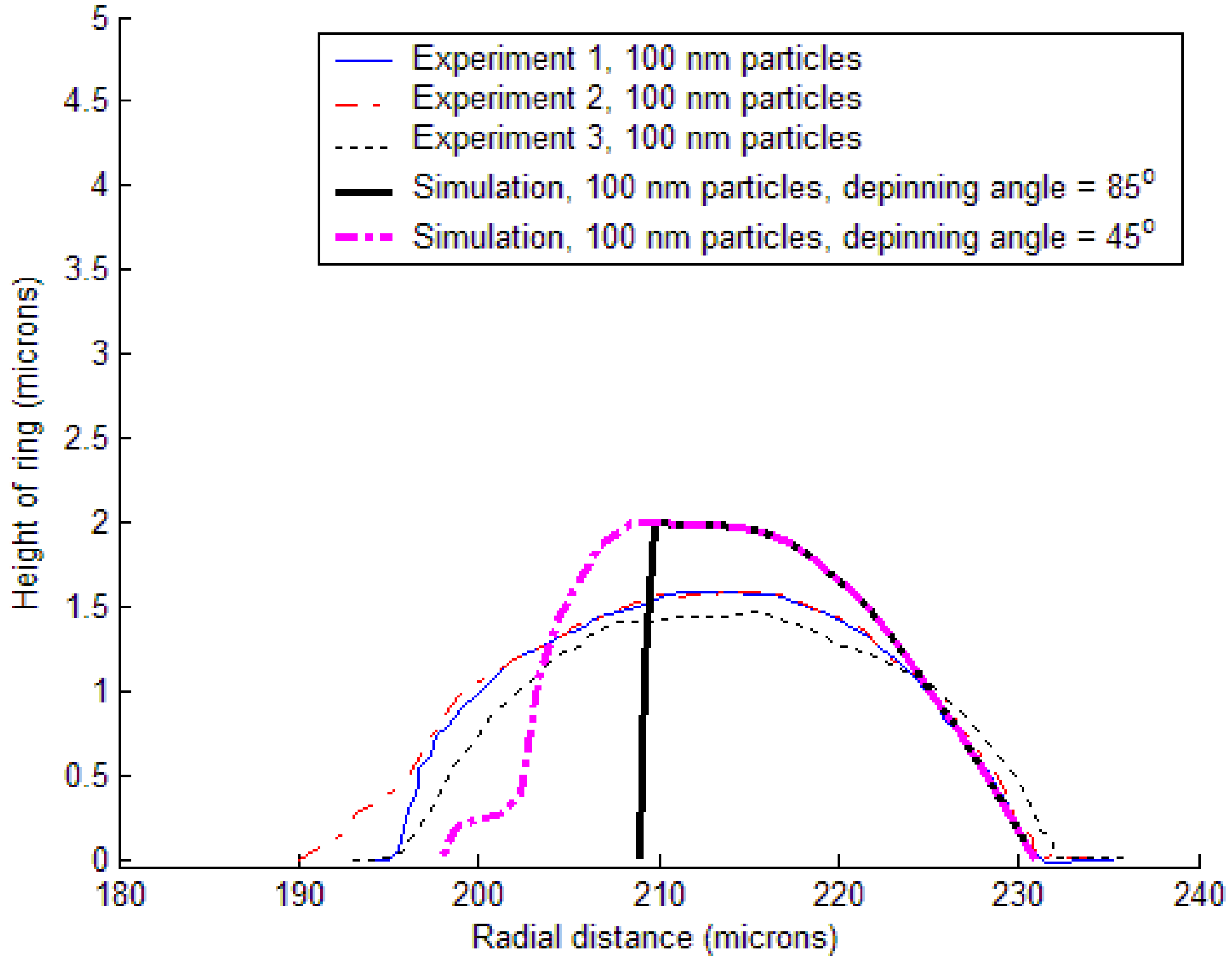

Figure 13: Comparison of AFM measurements and numerical results for the ring profile formed after the drying of a 3.7 nL water drop on glass substrate at ambient temperature for 100 nm particles. Initial concentration of the particles is 1%. Numerical results are shown for two values of the depinning angle ($\phi_{rec}$): 85° and 45°.

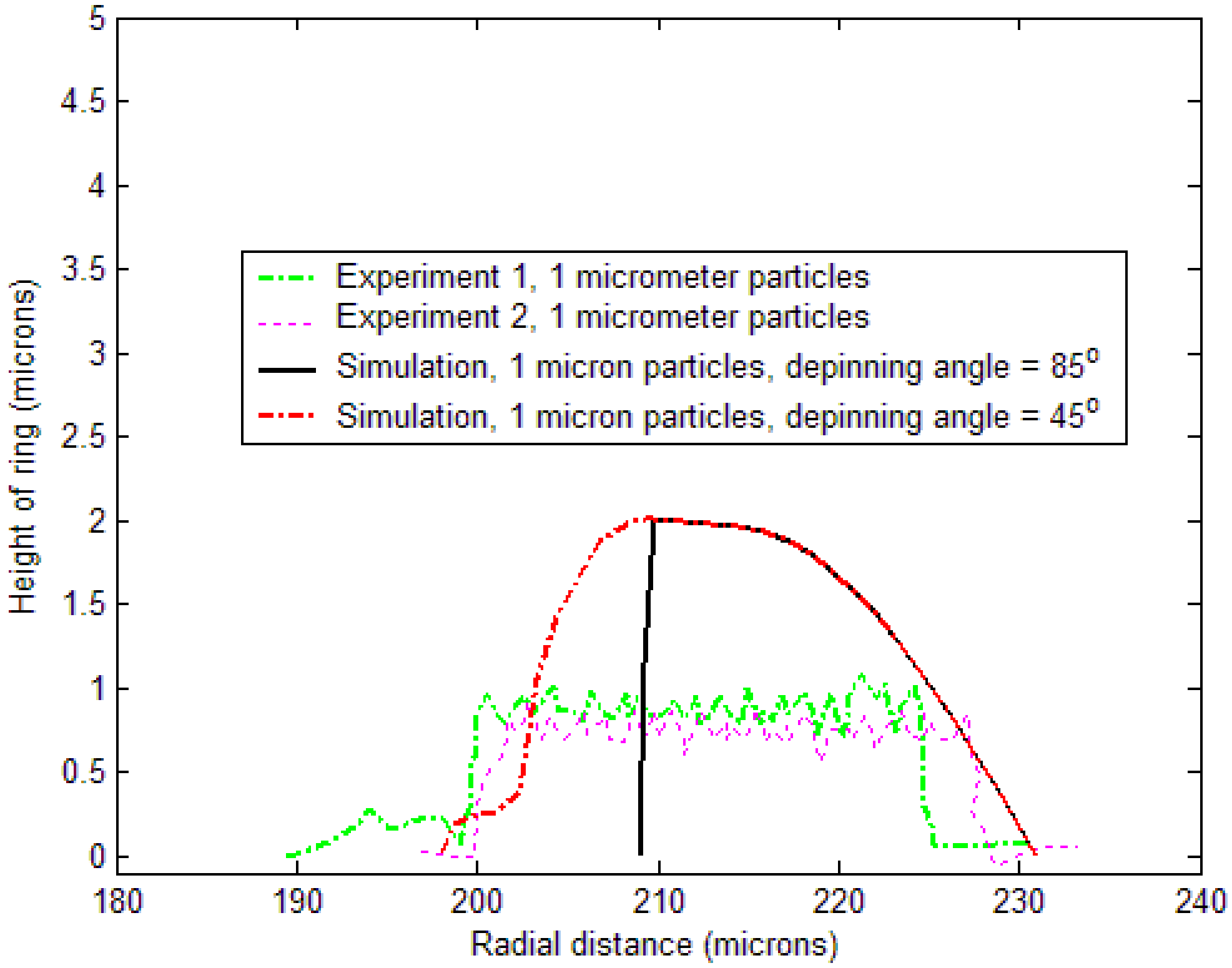

Figure 14: Comparison of AFM measurements and numerical results for the profile of the ring formed after the drying of a 3.7 nL water drop on glass substrate at ambient temperature for larger 1 μm particles. Initial concentration of the particles is 1%. Numerical results are shown for two values of the depinning angle ($\phi_{rec}$): 85° and 45°.

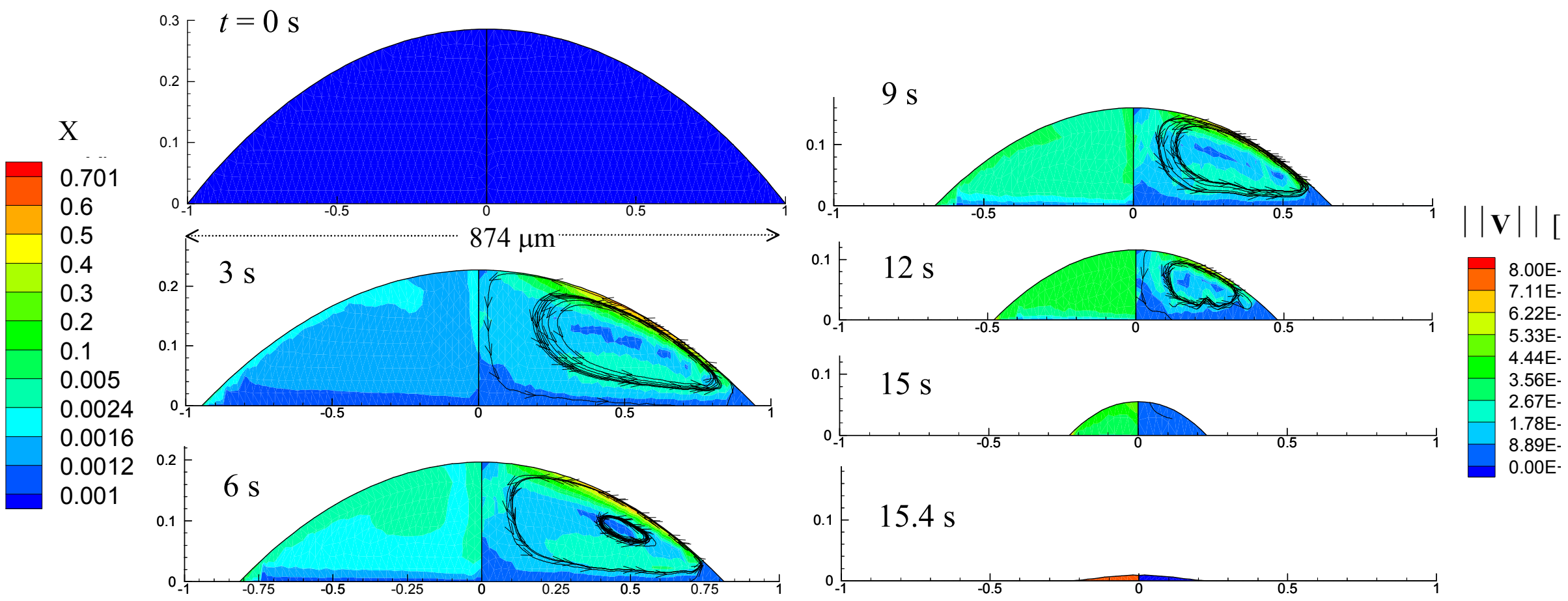

Figure 15: Simulation of the evaporation of a 38 nL colloidal isopropanol drop on PDMS for homogeneous bump formation. Particle concentration contours (left) and streamlines superposed to velocity amplitude (right) are shown. Note that the aspect ratio is increased towards the vertical axis from $t = 0$ to 12 s. See also associated movie [91].

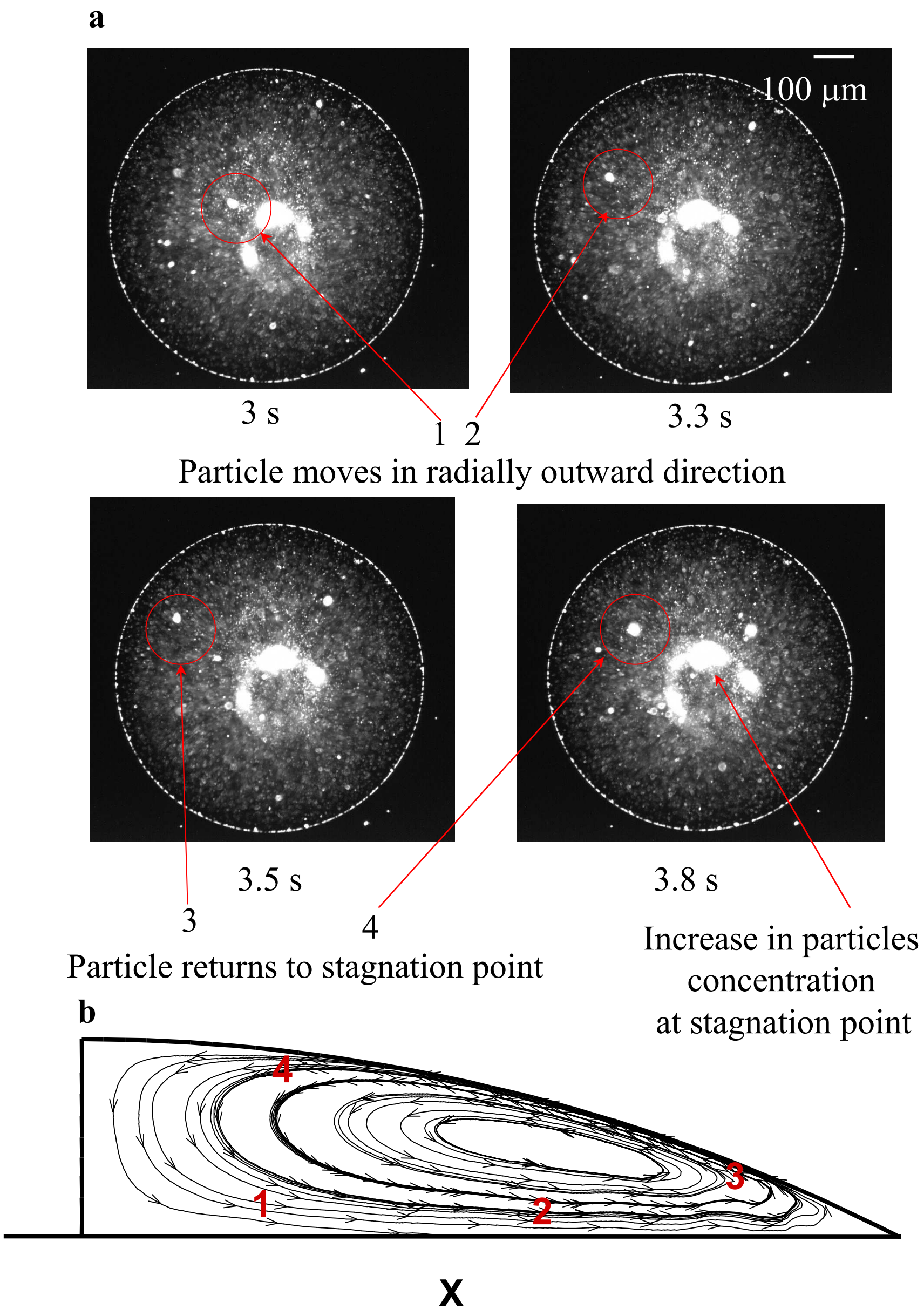

Figure 16: Particle visualization in the evaporating 38 nL isopropanol drop on PDMS using fluorescent microscopy. Initial substrate temperature, initial wetting angle and initial wetted radius of the drop are 25°C, 32° and 437 µm respectively. (a) A fluorescent particle is tracked at different positions 1, 2, 3 and 4 at different times (b) Corresponding positions of the particle in the drop as seen from the side of the drop. See also associated movie [86].

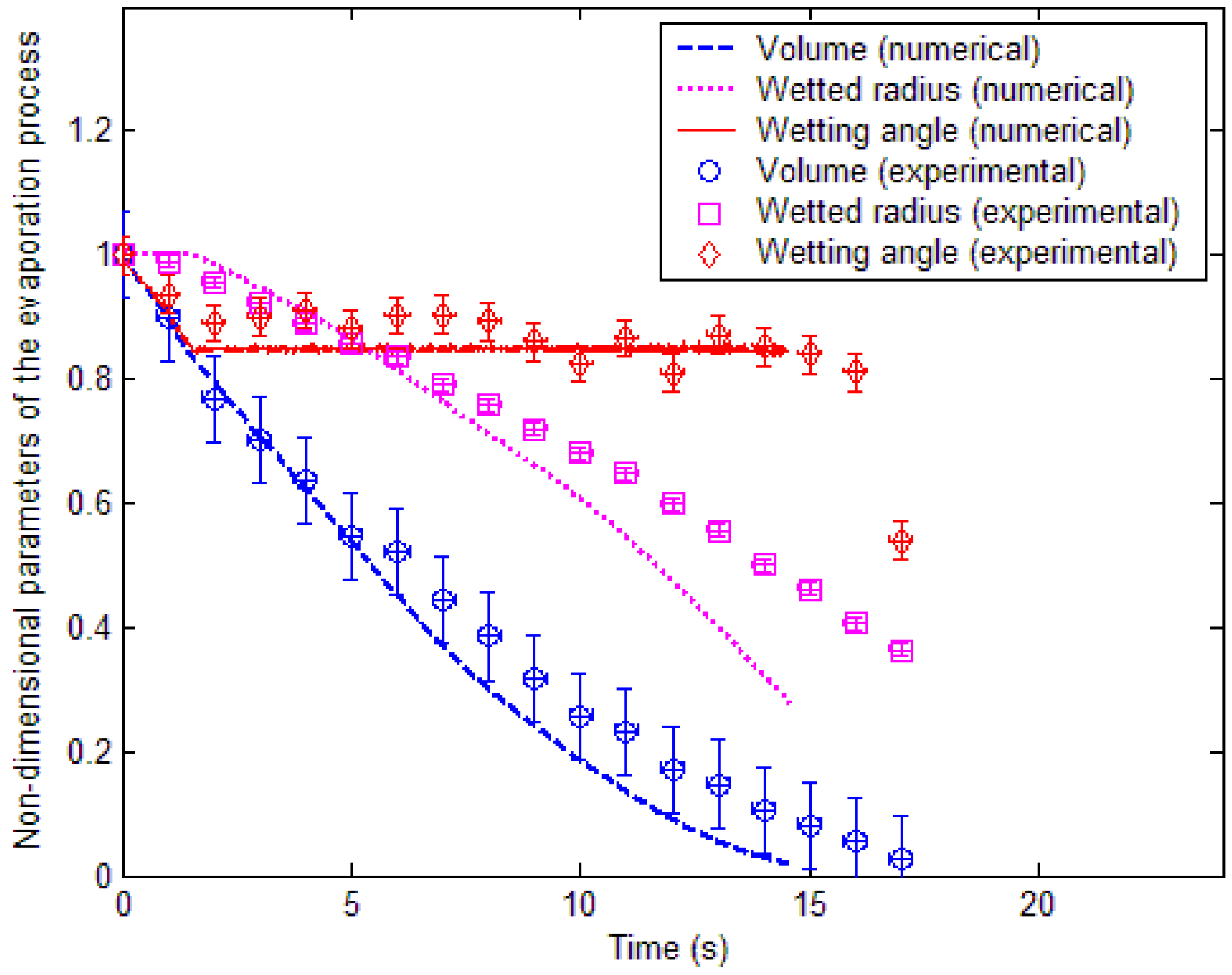

Figure 17: Time evolution of numerical and experimental volume, wetted radius and wetting angle for the evaporation of a 38 nL isopropanol drop on PDMS. Initial values of substrate temperature, wetting angle and wetted radius of drop are 25$^o$C, 32$^o$ and 437 μm respectively. Comparison of the evolution of volume, wetted radius and wetting angle is shown. These values are normalized with respect to their initial values (at $t = 0$). Error bars are also shown for experimental results.

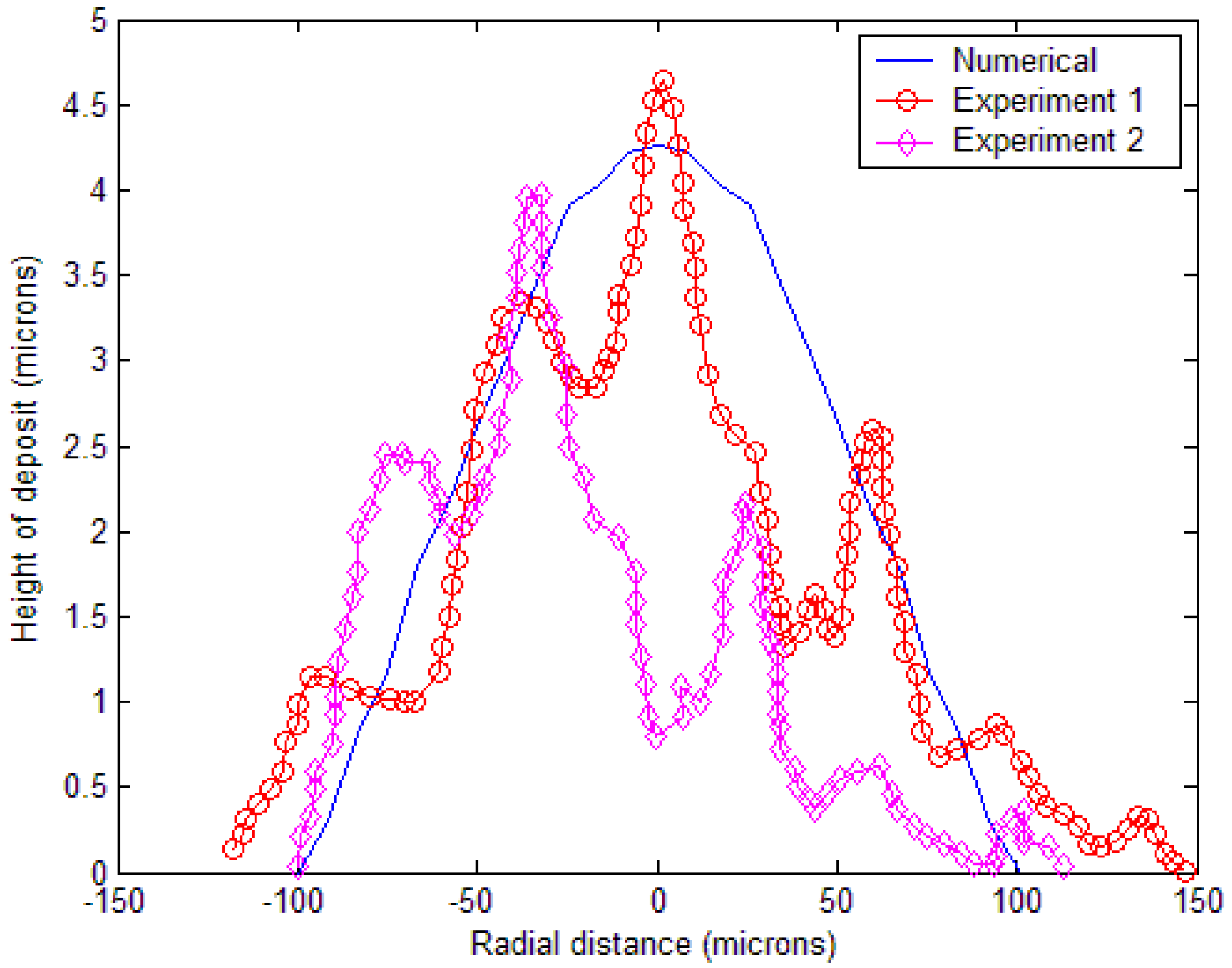

Figure 18: Numerical and experimental (using surface profilometer) results for the deposit profile obtained after the drying of a 38 nL isopropanol drop on PDMS at ambient temperature. Initial substrate temperature, initial wetting angle and initial wetted radius of drop are 25$^{o}$C, 32$^{o}$ and 437 μm respectively.